\documentclass[a4paper,11pt]{article}
\usepackage{amsfonts}
\usepackage{mathrsfs}
\usepackage{amsmath}
\usepackage{amssymb}
\usepackage[medium]{titlesec}
\usepackage{bm}
\usepackage{cite}
\usepackage{feynmp}
\usepackage[normalem]{ulem}
\usepackage{extarrows}
\usepackage{slashed}
\usepackage{isodateo}
\usepackage{graphicx}
\graphicspath{{fig/}}
\usepackage{xcolor}
\usepackage[bookmarksnumbered=true,bookmarksopen=true]{hyperref}
 \hypersetup{colorlinks,%
             linkcolor=[rgb]{0,0.2,0.6}, %
             citecolor=[rgb]{0,0.2,0.6}}
\usepackage{geometry}
\usepackage{indentfirst}
\newcommand{\FR}[2]{\displaystyle\frac{\,{#1}\,}{#2}}

\newcommand{\n}{\nonumber}

\def\bge{\begin{equation}}
\def\ede{\end{equation}}
\def\bga{\begin{aligned}}
\def\eda{\end{aligned}}
\def\bgp{\begin{pmatrix}}
\def\edp{\end{pmatrix}}
\def\bgs{\begin{subequations}}
\def\eds{\end{subequations}}
\newcommand{\order}[1]{\mathcal{O}({#1})}

\def\di{{\mathrm{d}}}
\def\D{{\mathrm{D}}}

\def\mb{\mathbf}

\def\pd{\partial}
\def\ld{{\mathscr{L}}}

\def\la{\langle}\def\ra{\rangle}
\def\sla{\slashed}

\def\tr{\mathrm{\,tr\,}}

\def\to{\rightarrow}
\def\To{\Rightarrow}
\def\ii{\mathrm{i}}

\def\al{\alpha}
\def\be{\beta}
\def\ga{\gamma}
\def\de{\delta}
\def\ep{\epsilon}

\def\lam{\lambda}
\def\rh{\rho}
\def\si{\sigma}

\def\Mp{M_{\text{Pl}}}

\def\Re{\mathrm{Re}\,}

\def\LIR{\Lambda_{\text{IR}}}

\def\eff{{\mathrm{eff}}}

\newcommand{\ob}[1]{\mkern 2mu \overline{\mkern -2mu #1 \mkern -2mu}\mkern 2mu}
\newcommand{\wt}[1]{\mkern 2mu \widetilde{\mkern -2mu #1 \mkern -2mu}\mkern 2mu}
\newcommand{\wh}[1]{\mkern 2mu \widehat{\mkern-2mu#1\mkern-2mu}\mkern 2mu}

\begin{document}

\title{\vspace{-5mm}\Large\textbf{Loop Corrections to Standard Model Fields in Inflation}}
\author{Xingang Chen$^{a,b}$\footnote{Email: Xingang.Chen@cfa.harvard.edu}{},~~ Yi Wang$^{c}$\footnote{Email: phyw@ust.hk}{},~ and~ Zhong-Zhi Xianyu$^{d}$\footnote{E-mail: xianyu@cmsa.fas.harvard.edu}\\[2mm]
\normalsize{$^a$\emph{Institute for Theory and Computation, Harvard-Smithsonian Center for Astrophysics,}}\\
\normalsize\emph{{60 Garden Street, Cambridge, MA 02138, USA}}\\
\normalsize{$^b$\emph{Department of Physics, The University of Texas at Dallas,}}\\
\normalsize{\emph{800 W Campbell Rd, Richardson, TX 75083, USA}}\\
\normalsize{$^c$\emph{Department of Physics, The Hong Kong University of Science and Technology,}}\\
\normalsize{\emph{Clear Water Bay, Kowloon, Hong Kong, P.R.China}}\\
\normalsize{$^d$\emph{Center of Mathematical Sciences and Applications, Harvard University,}} \\
\normalsize{\emph{20 Garden Street, Cambridge, MA 02138, USA}}}
\date{}

\maketitle

\begin{abstract}
We calculate 1-loop corrections to the Schwinger-Keldysh propagators of Standard-Model-like fields of spin-0, 1/2, and 1, with all renormalizable interactions during inflation. We pay special attention to the late-time divergences of loop corrections, and show that the divergences can be resummed into finite results in the late-time limit using dynamical renormalization group method. This is our first step toward studying both the Standard Model and new physics in the primordial universe.
\end{abstract}


\section{Introduction}

Cosmic Inflation has long been a leading paradigm describing the universe at its very early stage \cite{Guth:1980zm, Linde:1981mu, Albrecht:1982wi}. It not only explains various puzzles of the Big Bang cosmology, but also provides the primordial fluctuations for the universe which seeded the CMB anisotropy and large scale structures as we see today. Its predictions for these primordial fluctuations agree very well with current observations.

The inflation occurred at very high energy, during which the Hubble scale of the universe can be as high as $10^{14}$GeV, much higher than any artificial collider one can imagine. On top of that, the measurements of inflation observables through CMB or large scale structures have achieved impressive precision \cite{Ade:2013ydc}, and the precision will be further increased in the foreseeable future \cite{Baumann:2008aq, Matsumura:2013aja, Dore:2014cca, Munoz:2015eqa}. Based on this observation, it has been proposed recently that inflation can be used as a probe of the particle mass and spin spectra of new physics at high energies \cite{Chen:2009we,Chen:2009zp,AHM,Baumann:2011nk,Assassi:2012zq,Noumi:2012vr}, and for this reason it is dubbed the ``cosmological collider'' by Arkani-Hamed and Maldacena \cite{AHM}.

There has been a world of inflation models \cite{Chen:2010xka,Wang:2013eqj,Baumann:2014nda}. Their general predictions can be well consistent with observations, but they are very difficult to be discriminated. On the other hand, while new physics beyond Standard Model is generally expected to appear at some high energy scale, we are not sure what the new physics would be and at which scale it lies. Though our ignorance of inflation and new physics is vast, we do know that the Standard Model (SM) of particle physics is a very good description of, with a few exceptions, almost all microscopic phenomena up to TeV scale, and we do know that all SM fields should exist during inflation, too. Therefore, before trying to discover any new physics based on inflation, it is of vital importance to study the behavior of these well-known SM fields in an inflating universe and their possible observable consequences. Following Arkani-Hamed and Maldacena, we would also dub this program as ``a calibration of the cosmological collider''.

In this paper, we will make the first step toward this goal, by studying the behavior of SM fields during inflation, including spin-0 Higgs boson, spin-1/2 fermion and spin-1 gauge boson. We will not assume specific model for inflation, but it is important to distinguish two classes of inflation models, depending on whether the inflaton is a Higgs boson that triggers the electroweak symmetry breaking and also gives mass to all SM fermions at low energies. When the inflaton is (or has a component of) the electroweak Higgs field, a scenario known as Higgs inflation \cite{Bezru07,Bezru13}, all SM fermions and gauge bosons would acquire huge mass during inflation, due to the very large VEV of the Higgs field. On the other hand, if the inflaton is some scalar other than Higgs field, all SM field would remain massless at the tree level during inflation as long as Higgs doesn't develop large VEV.

In this paper we will focus on the non-Higgs inflation scenarios, and leave a parallel study of the very intriguing possibility of Higgs inflation to a future work. As mentioned above, in generic non-Higgs inflation scenarios all SM fields would apparently be massless, but this is only at the tree-level. Loop corrections can be very important for massless fields during inflation as they can generate late-time divergences. There have been extensive studies of such late-time divergences (sometimes known as the IR growth) of massless loop correction in inflation background as well as in de Sitter space (dS) and its Euclidean version. Similar IR issue is also known for a long time in thermal field theory.

In all these cases, the late-time divergences can be traced back to the existence of zero modes, which makes some higher order loop diagrams just as important as one loop diagrams. It is also known that once we properly resum these IR divergent loop diagrams, an IR finite result can be obtained, and in some cases the massless field would develop a nonzero mass due to the IR resummation.

We expect similar phenomenon appear for all SM fields during inflation. That is, the usual SM interactions among these fields, together with their tree-level zero mass, would generate a whole series of IR divergent loop correction to their two-point functions. The goal of this paper is to calculate these IR divergent loop corrections to two-point functions of SM fields and resum them consistently using Dynamical Renormalization Group method. As our first step, we will make the calculation for a general massless scalar boson, a general massless Dirac fermion, and a general gauge boson, respectively, assuming all renormalizable interactions among these fields, to mimic the SM dynamics. We will generalize the results in this paper to the realistic SM and study their observable consequences in a follow-up work \cite{prep}.

\begin{figure}[tbph]
\centering
\includegraphics[width=0.6\textwidth]{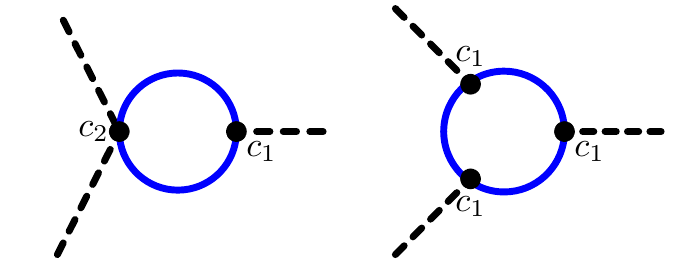}
\emph{
\caption{Two examples of 3-point function of inflaton field with SM contribution via 1-loop diagrams. The dashed black lines represent inflaton field, the solid blue lines represent SM fields, and $c_{1,2}$ denotes couplings between inflaton and SM fields.}
\label{fig_SMloop}
}
\end{figure}

Then it remains to understand how we can probe these quantum-corrected two-point functions through inflation observables. For this point we note that all SM fields carry some gauge charges so they enter non-Gaussianities only through loop diagrams. In the most pessimistic situation, the inflaton interacts with SM fields only through gravitation. But it is well possible that the interaction between inflaton and SM fields are much stronger. In fact if we assume the inflaton $\varphi$ comes from a sector of new physics whose energy scale is $M$, then in effective field theory it is expected that the inflaton $\varphi$ can interact with SM fields through following terms in the Lagrangian,
\bge
\Delta\ld=\Big(
   c_1\FR{\varphi}{M}
 + c_2\FR{\varphi^2}{M^2}
 + \cdots
 + \tilde c_3 \frac{\partial^2\varphi}{M^3}
 + \tilde c_4 \frac{(\partial\varphi)^2}{M^4}
 + \cdots
 \Big)\mathcal{O}_{\text{SM}},
\ede
where $\mathcal{O}_{\text{SM}}$ represents whatever gauge singlet operator formed by SM fields and the $c$-coefficients are typically $\order{1}$ coupling constants. When $M\ll \Mp$, this interaction will be greatly enhanced relative to gravitation, and thus make the loop signal of SM fields potentially detectable, through process like the diagrams in Fig.\,\ref{fig_SMloop}. In these diagrams, the dashed lines represent inflaton, and the blue lines represent any of SM fields. Here the blue lines have included quantum corrections with possible late-time divergences resummed. In this paper we will focus on the blue lines and leave the full study of this 3-point function to the next paper.

 Before entering the details of 1-loop calculation, we shall firstly review briefly the issue of late-time divergence in dS in Sec.\;\ref{sec2}. Then we set up the formalism and fix the notations for our calculation in Sec.\;\ref{sec3}. The explicit calculation of all 1-loop corrected in-in propagators is presented in Sec.\;\ref{sec4}. In Sec.\;\ref{sec5} we perform the resummation of late-time divergence of 1-loop propagators using dynamical renormalization group method. Further discussions about the current and future works are presented in Sec.\;\ref{sec6}. Readers not interested in details of calculation can skip Sec.\;\ref{sec4} and go directly to Sec.\;\ref{sec5} for a summary of our results.

\section{Loop Corrections and Late-Time Divergences}
\label{sec2}

The IR and late-time divergences of loop corrections in dS are well known and extensively studied. However, to the best of our knowledge, no common consensus is reached on a complete physical interpretation of them. In simple cases such as $\lam\phi^4$ theory with minimally coupled massless scalar field $\phi$ in dS, we do have various hints that the late-time divergences may be viewed as a signal of mass generation for the massless field $\phi$. However even this result may not be conclusive, and in more general situations, one can even find conflicting statements in the vast literature about this problem.

At first sight, the theoretical uncertainties in our understanding of late-time divergences may be an obstacle of phenomenological studies. But at the same time it also justifies the importance of such studies because one may hope to shed light on these theoretical issues by studying their potentially observable effects. Given the robustness of our understanding of SM physics, and the fast development of cosmological observation, we think it may be a good time to tackle this issue from phenomenological perspective.

Before making the first step toward this goal, in the current section we try to summarize what we know about the loop corrections and their late-time divergences in dS. We don't aim at a complete review of this issue, but only point out several known results relevant to our current study.

In the simplest case of a minimally coupled massless scalar field with $\lam\phi^4$ self-interaction in dS, it has long been known that the IR dynamics generates an effective mass $m_\eff$ for the scalar, with $m_\eff^2\propto \sqrt{\lam}H^2$, where $H$ is the Hubble scale. This result can be found by simple estimation using Hartree-Fock approximation, or more rigorous stochastic approach \cite{Linde82,Staro82,Staro94,Finelli:2008zg,Finelli:2010sh,Burge15}, or through 1-loop resummation using dynamical renormalization group method \cite{Burge09,Burge10} (See also \cite{Onemli02,Brunier05,Kahya10,Onemli14} for related diagrammatic calculations). On the other hand, in the Euclidean version of dS, it was recognized that the 1-loop late-time divergence is from the improper treatment of scalar zero mode. In that case the zero mode partition function is simple enough to be solved exactly, and the result shows that the zero mode acquire an effective mass $m_\eff^2\propto \sqrt{\lam}H^2$ \cite{Rajar10,Benek13}. This result can also be viewed as a supporting argument for massless scalar in dS because it was proved that the correlation functions calculated in Euclidean dS (namely a sphere) and in the Poincar\'e patch of dS using in-in formalism are equivalent\cite{Marolf1006,Marolf1010,Higuchi11} (see however \cite{Akhme11,Krotov10,Polya12} for different perspectives).

The appearance of an effective mass for the classically massless scalar field due to IR dynamics is reminiscent of what happens in ordinary flat-space thermal field theory, in which case the IR dynamics of massless $\lam\phi^4$ theory will also introduce an effective mass $m_\eff^2=\lam T^2$ to the zero mode of $\phi$ (See, e.g., \cite{Kapusta}). Meanwhile it is also known that an observer moving along timelike geodesic in dS would find Bunch-Davis vacuum to be a thermal state with temperature $T=H/2\pi$. So one may attempt to apply this result directly and conclude that the effective mass in dS's case is $m_\eff^2\propto \lam H^2$. But this is not correct. Because the value of the frozen super-Hubble fluctuations are determined by the competition between the random walk behavior and the classical roll-down behavior on its potential. The random walk behaves as $\phi^2\sim H^3t$ and the classical slow roll equation of motion can be solved as $\phi^2 \sim H/(\lambda t)$. The balance takes place at $t\sim 1/(H\sqrt\lambda)$, where $\phi^2\sim H^2/\sqrt{\lambda}$ and $m_\mathrm{eff}^2 \sim \lambda\phi^2\sim \sqrt{\lambda}H^2$. This is different from the flat-space thermal behavior where $\phi\sim T$ do not suffer from the IR raldom-walk type of growth. 

More fundamentally, the thermal nature of ordinary finite-temperature field theory in flat spacetime is quite different from dS. For instance, in former's case, the theory can be thought of as a Wick-rotated Euclidean space $\mathbb{R}^3\times S^1$ with imaginary time compactified to a circle of radius $(2\pi T)^{-1}$, and this manipulation breaks the isometry of spacetime (Lorentz transformation in particular). As a result, there is a preferred frame of reference, and initial observers moving relative to this frame will see the anisotropic thermal radiation due to Doppler shift. However, in dS, if we try to use Wick rotation to study its thermal nature, then the Wick-rotated space is the sphere $S^4$, and the thermal state (Bunch-Davis) in this case doesn't break any of isometry of dS. This implies in particular that the Bunch-Davis vacuum appears to have the same temperature to observers moving along any timelike geodesic in dS. Now, the effective mass generated for $\phi$ in both cases are from zero modes, and it is obvious that the zero modes in $\mathbb{R}^3\times S^1$ and in $S^4$ are quite different, so it is conceivable that they contribute to effective mass differently.

In fact the thermal nature as well as IR problem of dS is much more obscure, partly because we are not sure whether it is correct to use the Wick-rotated sphere to understand all puzzles associated with late-time divergences. Indeed, in direct calculation using real time Schwinger-Keldysh formalism \cite{SKF1,SKF2}, the late-time divergences of a field theory may or may not be interpreted as the signal of mass generation. In some cases as we shall show in this paper, the 1-loop divergences cannot even be subtracted by local counterterms, and do not resemble an effective mass term at all. But on the other hand, in $S^4$ there cannot be late-time divergences at all because the space itself is bounded and the radius of the sphere provides a natural IR cutoff to the calculation. The prove of equivalence between correlation functions on sphere and in the Poincar\'e patch in \cite{Marolf1010,Higuchi11} seems to apply only to massive scalar fields, and it is less clear if we can trust this equivalence for classically massless scalar field. Furthermore, as argued in \cite{Krotov10,Polya12}, the finite temperature of Bunch-Davis vacuum as derived by performing Wick rotation to sphere actually describes the situation that all thermal radiation emitted out of horizon is somehow artificially reflected back, and thus cannot capture the instability due to the particle production, nor the late-time divergences generated by loop modes outside the horizon. This is much like the fact that the Euclidean version of Rindler space cannot capture the evaporation of Schwarzschild black hole in asymptotically flat spacetime, and an artificially introduced thermal bath is needed to maintain the thermal equilibrium.

Beyond the simplest $\lam\phi^4$ theory, the situation is even more unclear. For instance, the 1-loop correction to scalar's two-point function from $\phi^3$ interaction is more nonlocal than a mass term, and so cannot be interpreted as mass generation at all \cite{Meulen07,Burge09,Burge10}. There are also studies of 1-loop correction to fields with spin other than 0. Relevant to our interest are the cases of fermions and vector bosons interacting with massless scalar field through Yukawa and gauge couplings, respectively.

For massless fermions, the Dirac mass term is forbidden by chiral symmetry, therefore even if the IR correction will generate some mass-like terms for massless fermion, these terms cannot be Dirac mass. This simple argument was confirmed by explicit calculation \cite{Garbr06}, and it was found there that the fermion indeed acquire a nonlocal mass-like term in its effective action from 1-loop correction of scalar field.

For vector boson, naively the mass term is forbidden by gauge symmetry. However the true story is more complicated. In flat-space thermal field theory, it is well known that the temporal component of photon can acquire a nonzero thermal mass which is responsible for Debye screening. This mass term is consistent with gauge invariance because the finite temperature breaks the Lorentz symmetry as mentioned above. As a result, the Ward identity will naturally admit the temporal component of gauge field to develop a nonzero mass (i.e. the so-called electric mass), while the mass of spatial components (so-called magnetic mass) is still forbidden. But the Bunch-Davis vacuum respects all isometries of dS, so the mass term for gauge field is still forbidden\footnote{However this argument would not apply if dS isometry is spontaneously broken by IR dynamics.}. This fact can also be understood with the picture of Wick-rotated space. In flat-space's case $\mathbb{R}^3\times S^1$, we do have a nontrivial gauge invariant for flat gauge field (By flat gauge field we mean the corresponding field strength is zero), which is the nontrivial gauge-invariant Wilson loop $e^{\ii\oint\di \tau A_0}$ surrounding the compact imaginary time direction. Therefore a gauge-invariant mass for $A^0$ can indeed be generated after Kaluza-Klein reduction. On the other hand, the Euclidean version of dS (namely $S^4$) is simply connected, so any gauge-invariant Wilson loop is trivial, and therefore the photon mass is still forbidden by gauge symmetry in dS, even there appears to be a nonzero temperature in Bunch-Davis vacuum. We can also understand this result using a less rigorous but very intuitive picture, in which we can think of the mass of photon in ordinary thermal field being generated by constant ``kicks'' of thermal particles as the photon travel through the thermal background. On the other hand, the Bunch-Davis vacuum can be viewed as a true vacuum without any thermal excitation by an lightlike geodesic observer (See, e.g., \cite{Greene05}), so the photon in Bunch-Davis vacuum won't experience the ``thermal kicks'' and thus will not receive corresponding thermal mass.

A possible photon mass generated from a massless scalar field in scalar QED was previously estimated to be $m_A^2\propto e^2H^2$ using Hartree-Fock approximation \cite{Davis00}. Such IR dynamics in scalar QED was then studied in great details \cite{Proko0205,Proko0210,Proko03,Proko04}, and it was found that the 1-loop correction from massless scalar field to photon's self-energy looks like a mass contribution. However it would be premature to interpret this result as a genuine photon mass because the mass obtained in \cite{Proko0205,Proko0210,Proko03,Proko04} is wave-number dependent while the genuine massive vector field cannot have wave-number dependent mass. Actually the 1-loop self-energy obtained in \cite{Proko0205,Proko0210,Proko03,Proko04} is nonlocal (conceivable in dS). So this result does not contradict the argument in last paragraph.

All results mentioned above considered loop corrections from a minimally coupled massless scalar field. As we will show below, this is the only possibility for the generation of late-time divergences \footnote{There have also been a lot of studies on graviton loops \cite{Tsamis:1996qm, Tsamis:1996qk, Dimastro2008, Bartolo2010, Xue:2012wi, Giddings:2010nc} in the literature. It is indeed possible that graviton loops can generate additional late time divergences because the tensor components of the metric also froze on super-Hubble scales and have random walk behavior. However, as we are considering direct couplings between the inflaton and the Standard Model sector, the graviton interaction is relatively much weaker and can be ignored for our setup.}. Massless fermions and massless gauge bosons cannot contribute such IR divergent loops essentially because their kinetic action as well as interactions are classically Weyl invariant. This can be understood by direct power counting of scale factor $a(\tau)$, as will be elaborated in Sec.\;\ref{sec4}. 

Phenomenologically, we are interested in the inflationary fluctuations instead of pure dS background. For the inflationary background, there are in general two kinds of fields, the curvature field and the iso-curvature fields. The fields that we mentioned above and we will be interested in this paper are the iso-curvature fields, namely some spectator fields in the inflation background. On the other hand, the inflaton field, which rolls down the inflationary potential and drive inflation, breaks the dS symmetry and its fluctuation behave differently from the above dS space results. This curvature perturbation (if not interacting with iso-curvature perturbations) is conserved on super-Hubble scales for attractor single field inflation models\footnote{Non-attractor inflation models introduce another type of super-Hubble evolution for the curvature mode without considering self-interactions \cite{Kinney:2005vj,Chen:2013aj,Namjoo:2012aa,Huang:2013lda,Chen:2013kta,Chen:2013eea}.} even when self-coupling is taken into account. This can be understood from the separate universe picture \cite{Lyth:2004gb}, operator product expansion and Goldstone theorem \cite{Assassi:2012et} or explicit calculations \cite{Senatore:2009cf,Senatore:2012nq,Pimentel:2012tw,Senatore:2012ya}.

Given the complication discussed in this section, we shall adopt in this paper a more practical approach, by simply calculating the 1-loop divergences with real time Schwinger-Keldysh formalism, without running into the complication of Wick rotation. We shall also resum the possible late-time divergences using dynamical renormalization group method. When the results resemble mass correction, we shall interpret them accordingly, otherwise we shall just leave them as they are. In any case, these results can be directly applied to the realistic SM fields, so that their imprints on primordial non-Gaussianities can be readily worked out.

\section{Standard-Model-Like Fields in Inflation Background}
\label{sec3}


\subsection{Schwinger-Keldysh Formalism}

The task of the current paper is to calculate the quantum corrections to in-in propagator of massless field with spin-0, 1/2, or 1 in dS background. Here and in the following by dS we always mean its Poincar\'e patch, parametrized by the conformal coordinates $(\tau,\mb x)$ with metric $\di s^2=a^2(\tau)(-\di\tau^2+\di\mb x^2)$, where $a(\tau)=-1/(H\tau)$ is the scale factor, and by in-in propagator we mean the expectation value,
\bge
  \Big\la\text{in}\Big|\phi(\tau,\mb k)\phi(\tau',\mb k')\Big|\text{in}\Big\ra,
\ede
where $\phi$ is an arbitrary field we are interested in, $\mb k$ is the Fourier transform variable of $d$-dimensional flat coordinates $\mb x$, and $|\text{in}\ra$ is the in-vacuum, which we will take to be Bunch-Davis vacuum.

It is convenient to use the diagrammatic approach to compute loop corrections to the in-in propagator. To this end we use the Schwinger-Keldysh formalism \cite{SKF1,SKF2}, in which the numbers of fields are doubled in order to take account of the closed time path in the in-in formalism. Then the in-in expectation values can be found by taking functional derivatives of the following generating functional $W[J_+,J_-]$:
\bge
  e^{\ii W[J_+,J_-]}=\int\mathcal{D}\phi_f(\tau_f)\int\mathcal{D}\phi_-\,e^{-\ii S[\phi_-]-\ii\int \di^4x\,J_-(x)\phi_-(x)}\int\mathcal{D}\phi_+\,\,e^{\ii S[\phi_+]+\ii\int \di^4x\,J_+(x)\phi_+(x)},
\ede
where $J_+$ and $J_-$ are classical sources associated with $\phi_+$ and $\phi_-$, respectively, $\tau_f$ is any time to the future of all times we are interested, and $\phi_f(\tau_f)$ is the field evaluated at $\tau_f$. The form of this generating functional is easy to understand: the $\phi_+$ and $\phi_-$ integral represent the generating functionals of ordinary in-out amplitudes and their conjugates, respectively, and the additional integral over $\phi_f(\tau_f)$ is simply to sum over the artificially introduced out states, making use of the completeness condition on Cauchy surface at $\tau_f$. Therefore the above generating functional should be subject to additional boundary condition $\phi_+(\tau_f)=\phi_-(\tau_f)=\phi_f(\tau_f).$

Then we also have four types of propagators:
\bge
  G_{ab}(x,y)=\ii\FR{\de}{a\ii\de J_a(x)}\FR{\de}{b\ii\de J_b(y)}e^{\ii W[J_+,J_-]}\bigg|_{J_+=J_-=0},
\ede
where $a,b$ represent either $+$ or $-$. Then it can be readily derived for scalar field $\phi$ that,
\bge
  \begin{pmatrix} G_{++}(x,y) & G_{+-}(x,y) \\ G_{-+}(x,y) & G_{--}(x,y) \end{pmatrix}
  =\ii\begin{pmatrix}
  \big\la \text{T}\phi(x)\phi(y) \big\ra &
  \big\la \phi(y)\phi(x) \big\ra \\
  \big\la \phi(x)\phi(y) \big\ra &
  \big\la \ob{\text{T}}\phi(x)\phi(y) \big\ra
  \end{pmatrix},
\ede
where in the expectation values T means time-ordered product and $\ob{\text{T}}$ means anti-time ordered product.

It will be convenient to work in momentum space, in which the real scalar field can be quantized as usual,
\bge
  \phi(\tau,\mb x)=\int\FR{\di^3k}{(2\pi)^3}\Big[u(\tau,\mb k)a(\mb k)+u^*(-\mb k)a^\dag(-\mb k)\Big]e^{\ii\mb k\cdot\mb x},
\ede
where $u(\tau,\mb k)$ is the mode function, and $a$, $a^\dag$ are corresponding annihilation and creation operators, respectively. Then the propagators above can be worked out more explicitly at tree level as,
\begin{align}
 \wh G_{++}(\mb k,\tau_1,\tau_2)
 =&~\ii \Big[\theta(\tau_1-\tau_2)u(\tau_1,\mb k)u^*(\tau_2,\mb k)+\theta(\tau_2-\tau_1)u(\tau_2,-\mb k)u^*(\tau_1,-\mb k)\Big],\\
 \wh G_{+-}(\mb k,\tau_1,\tau_2)
 =&~ \ii u(\tau_2,-\mb k)u^*(\tau_1,-\mb k),\\
 \wh G_{-+}(\mb k,\tau_1,\tau_2)
 =&~ \ii u(\tau_1,\mb k)u^*(\tau_2,\mb k),\label{ScaMPP}\\
 \wh G_{--}(\mb k,\tau_1,\tau_2) 
 =&~\ii \Big[\theta(\tau_1-\tau_2)u(\tau_2,-\mb k)u^*(\tau_1,-\mb k)+\theta(\tau_2-\tau_1)u(\tau_1,\mb k)u^*(\tau_2,\mb k)\Big],
\end{align}
where the hat on $\wh G$ means that the $\delta$-function factor $(2\pi)^3\de^{(3)}(\mb k+\mb k')$ has been amputated. In the rest of this section and in Sec.\;\ref{sec4}, whenever we mention the propagator $\wh{G}$ we always means its tree-level part, unless otherwise stated.

In the same way, all vertices, including internal interaction vertices and external vertices for external fields, also come in two types, $+$ or $-$, with an additional sign of $+$ or $-$ associated in the Feynman rules.

At the end of this subsection, we calculate the mode function $u(\tau,\mb k)$ of a scalar field $\phi$ with mass $m$ and minimally coupled to gravity, which is the positive-frequency solution to its equation of motion. In $(d+1)$-dimensional dS, the equation of motion reads,
\bge
  \phi_{\mb k}''(\tau)-\FR{d-1}{\tau}\phi_{\mb k}'(\tau)+k^2\phi_{\mb k}(\tau)+\FR{m^2}{H^2\tau^2}\phi_{\mb k}(\tau)=0.
\ede
We solve this equation to get the mode function,
\bge
\label{MSM}
  u(\tau,\mb k)=-\ii\FR{\sqrt\pi}{2}e^{\ii \pi(\nu/2+1/4)}H^{(d-1)/2}(-\tau)^{d/2}\mathrm{H}_{\nu}^{(1)}(-k\tau),
\ede
where $\nu\equiv\sqrt{\frac{d^2}{4}-\frac{m^2}{H^2}}$, $\mathrm{H}_{\nu}^{(1)}(z)$ denotes the Hankel function of first kind, and the normalization of $u$ is fixed by the inner product $a^{d-1}(\tau)(u\dot u^*-\dot uu^*)=\ii$ up to the phase (Here $\dot u\equiv \di u/\di\tau$). In the case of $d=3$ and $m=0$, the mode function reduces to the familiar one,
\bge
\label{MLSM}
  u(\tau,\mb k)=\FR{H}{\sqrt{2k^3}}(1+\ii k\tau)e^{-\ii k\tau}.
\ede

\subsection{Dirac Fermion}

The classical action of a massless Dirac Fermion $\psi$ field in general $(d+1)$-dimensional curved background can be written as,
\bge
  S_f=\FR{\ii}{2}\int\di^{d+1}x\,\det(e_\nu^n)\Big(\ob{\psi}\ga^\mu\nabla_\mu\psi-\ob\psi\overleftarrow\nabla_\mu\gamma^\mu\psi\Big),
\ede
where $e_\nu^n$ is the vierbein, $\gamma^\mu=\gamma^m e_m^\mu$ is the gamma matrices. Here greek indices correspond to curved coordinates and latin indices correspond to local Lorentz frame. The covariant derivative for Dirac spinor is $\nabla_\mu\psi=(\pd_\mu+\frac{1}{4}\omega_\mu{}^{mn}\si_{mn})\psi$, where $\omega_\mu{}^{mn}$ is the spin connection and $\si_{mn}=-\frac{1}{2}[\ga_m,\ga_n]$. In the conformal coordinates we are working with, it is easy to calculate the metric $g_{\mu\nu}=(H\tau)^{-2}\eta_{\mu\nu}$, the vierbein $e_\mu^m=(-H\tau)^{-1}\de_\mu^m$, and the spin connection $\omega_\mu{}^{mn}=(-\tau)^{-1}(\de_\mu^m\de_0^n-\de_\mu^n\de_0^m)$. Then the Dirac equation $\ii\gamma^\mu\nabla_\mu\psi=0$ derived from $\de S_f/\de\ob\psi=0$ gets simplified to $\ii(\sla\pd-\frac{d}{2\tau}\ga^0)\psi=0$.

Now it's easy to see that the solution to the Dirac equation can be written as $\psi=(-H\tau)^{d/2}\wt\psi$, with $\wt\psi$ the solution to the corresponding Dirac equation $\ii\sla\pd\wt\psi=0$ in flat spacetime. This allows us to get the mode function for $\psi$ by directly rescaling the corresponding flat space result, that is,
\begin{align}
&\psi(x)=(-H\tau)^{d/2}\int\FR{\di^dk}{(2\pi)^d}\sum_s\Big[\al_s(\mb k)a_s(\mb k)e^{-\ii k^0t}+\be_s(-\mb k)b^\dag_s(-\mb k)e^{+\ii k^0t}\Big]e^{+\ii\mb k\cdot\mb x},\\
&\ob{\psi}(x)=(-H\tau)^{d/2}\int\FR{\di^dk}{(2\pi)^d}\sum_s\Big[\ob\be_s(\mb k)b_s(\mb k)e^{-\ii k^0t}+\ob\al_s(-\mb k)a^\dag_s(-\mb k)e^{+\ii k^0t}\Big]e^{+\ii\mb k\cdot\mb x},
\end{align}
where $k\equiv|\mb k|$, $\al(\mb k)$ and $\be(\mb k)$ are the Dirac spinors with definite conformal momentum $\mb k$, and the creation and annihilation operators satisfy the usual anticommutation relation, with the nonzero ones given by,
\bge
\{a_s(\mb k),a_{s'}^\dag(-\mb k')\}=\{b_s(\mb k),b_{s'}^\dag(-\mb k')\}=(2\pi)^d\de_{ss'}\de^{(d)}(\mb k+\mb k').
\ede
We note that the mode function for $\psi$ is a simple rescaling of the corresponding flat spacetime result, because the action $S_f$ is Weyl invariant. This will not be the case if $\psi$ is massive. To see this, we add the mass term to the above action,
\bge
  S_f(m)=\int\di^{d+1}x\,\det(e_\nu^n)\bigg[\FR{\ii}{2}\Big(\ob{\psi}\ga^\mu\nabla_\mu\psi-\ob\psi\overleftarrow\nabla_\mu\gamma^\mu\psi\Big)-m\ob\psi\psi\bigg].
\ede
Then the equation of motion will be $(\ii\gamma^\mu\nabla_\mu-m)\psi=0$. Now if we still rescale the field as $\psi=(-H\tau)^{d/2}\wt\psi$, then the rescaled field $\wt\psi$ will obey the equation $\big[\ii\gamma^m\pd_m+m/(H\tau)\big]\psi=0$ which no longer resembles its flat-space counterpart.

The Schwinger-Keldysh formalism can be readily applied to fermions, for which one only needs to be careful with ordering since spinors are represented by Grassmann variables in the path integral,
\begin{align}
  Z[I_-,\ob I_-;I_+,\ob I_+]\equiv&~\int\mathcal{D}\ob{\psi}_f(\tau_f)\mathcal{D}{\psi}_f(\tau_f)\int\mathcal{D}\ob{\psi}_-\mathcal{D}{\psi}_- \,e^{-\ii S[\psi_-,\ob{\psi}_-]-\ii \int\di^4x\,(\ob I_-\psi_-+\ob \psi_- I_-)}\n\\
  &~\times\int\mathcal{D}\ob{\psi}_+\mathcal{D}{\psi}_+ \,e^{\ii S[\psi_+,\ob{\psi}_+]+\ii\int\di^4x\,( \ob I_+\psi_+ +\ob \psi_+ I_+)},
\end{align}
where $I_{\pm}$ and $\ob{I}_\pm$ are corresponding classical sources. For a massless Dirac fermion field $\psi$, we can also derive its various propagators, defined via,
\bge
  G_{ab}^{(F)}(x,x')=\ii\FR{\de}{a\ii\de\ob I_a(x)}\FR{\de}{ -b\ii \de I_b(x')}Z[I_-,\ob I_-;I_+,\ob I_+]\bigg|_{\text{all~} I\text{'s}=0},
\ede
where $a,b$ represent either $+$ or $-$.
Then in momentum space, we can derive,
\bge
\begin{pmatrix}
  {G}_{++}^{(F)}(\mb k,\tau;\mb k',\tau')
 &{G}_{+-}^{(F)}(\mb k,\tau;\mb k',\tau')\\
  {G}_{-+}^{(F)}(\mb k,\tau;\mb k',\tau')
 &{G}_{--}^{(F)}(\mb k,\tau;\mb k',\tau')
\end{pmatrix}
  =
\ii\begin{pmatrix}
  \la \text{T} \psi(\tau,\mb k)\ob{\psi}(\tau',\mb k')\ra
 &-\la \ob\psi(\tau',\mb k')\psi(\tau,\mb k)\ra\\
  \la \psi(\tau,\mb k)\ob{\psi}(\tau',\mb k')\ra
 &-\la \ob{\text{T}} \ob{\psi}(\tau',\mb k'){\psi}(\tau,\mb k)\ra
\end{pmatrix}
\ede
More explicitly, we can work out the tree propagators, with the results written in amputated form $ {G}_{ab}^{(F)}(\mb k,\tau;\mb k',\tau')= \wh{G}_{ab}^{(F)}(\mb k,\tau,\tau')(2\pi)^d\de^{(d)}(\mb k+\mb k')$. For example,
\begin{align}
 {\wh{G}}_{++}^{(F)}(\mb k,\tau,\tau')
 =&~\ii(H^2\tau\tau')^{d/2}\sum_s\Big[\theta(\tau-\tau')\al_s(\mb k)\ob\al_s(\mb k)e^{-\ii k(\tau-\tau')}\n\\
  &~~~~~~~~~~~~~~~~~~~-\theta(\tau'-\tau)\be_s(-\mb k)\ob\be_s(-\mb k)e^{-\ii k(\tau'-\tau)}\Big]\n\\
 =&- \FR{\ii(H^2\tau\tau')^{d/2}}{2k}\Big[\theta(\tau-\tau')\sla k e^{-\ii k(\tau-\tau')}-\theta(\tau'-\tau)\ob{\sla{k}}e^{-\ii k(\tau'-\tau)}\Big],
\end{align}
where we have used the spin sum relation $\sum_s\al_s(\mb k)\ob\al_s(\mb k)=-\sla k/(2k)$ and $\sum_s\be_s(\mb k)\ob\be_s(\mb k)=-\sla k/(2k)$. Here $\sla k\equiv \ga^\mu k_\mu$ and $\ob{k}^\mu=(k^0,-\mb k)$. The contracted indices here are all flat Lorentz indices. Similarly, we have,
\begin{align}
  &\wh G_{+-}^{(F)}(\mb k,\tau,\tau')=+\FR{\ii(H^2\tau\tau')^{d/2}}{2k}\ob{\sla k} e^{-\ii k(\tau'-\tau)},\\
  &\wh G_{-+}^{(F)}(\mb k,\tau,\tau')=-\FR{\ii(H^2\tau\tau')^{d/2}}{2k}\sla k e^{-\ii k(\tau-\tau')},\label{FerMPP}\\
  &\wh G_{--}^{(F)}(\mb k,\tau,\tau')=+\FR{\ii(H^2\tau\tau')^{d/2}}{2k}\Big[\theta(\tau-\tau')\ob{\sla{k}}e^{-\ii k(\tau'-\tau)}-\theta(\tau'-\tau)\sla k e^{-\ii k(\tau-\tau')}\Big].
\end{align}

\subsection{Photon}

Both Abelian and non-Abelian gauge fields are present in SM. For non-Abelian gauge field, its self interaction cannot generate any late-time divergence (cf. Sec.\;\ref{sec4}). So we shall only consider Abelian gauge field in the current work, and the corresponding results for non-Abelian gauge fields can simply be got by including appropriate group factors. 

For Abelian gauge field (which we also call photon for simplicity), the kinetic action is explicitly Weyl invariant only in 4-dimensional spacetime. So let's firstly consider this much simpler case, and put a general discussion on $(d+1)$-dimensional case to the end of this subsection. In 4 dimensions, the Weyl invariant action of photon is given by,
\bge
  S_V=-\FR{1}{4}\int\di^4x\,\sqrt{-g}g^{\mu\rh}g^{\nu\si}F_{\mu\nu}F_{\rh\si}.
\ede
As a consequence, the action, the equation of motion, and the mode functions for photon are the same with the corresponding ones in flat spacetime, once the conformal coordinates are chosen.

It is straightforward to repeat the standard path integral quantization of gauge field in flat space \`a la Faddeev-Popov. One  complication here is that the usual Lorentz gauge fixing term with arbitrary gauge parameter $\zeta$ is not Weyl invariant,
\bge
  S_{GF}(\text{Lorentz})=-\FR{1}{2\zeta}\int\di^4x\,\sqrt{-g}(g^{\mu\nu}\nabla_\mu A_\nu)^2,
\ede
which introduces additional $\tau$-dependent terms in to the classical action. So we solve this problem by choosing the following gauge:
\bge
  S_{GF}=-\FR{1}{2\zeta}\int\di^4x\,\sqrt{-g}\Big(g^{\mu\nu}\nabla_\mu A_\nu+\FR{2}{\tau}A_0\Big)^2.
\ede
That is, we choose the gauge in such a way that the quantized action of the gauge field still resembles its flat-space counterpart, at the expense of losing the manifest general covariance. Now it is easy to show that after choosing the conformal coordinates in dS, the quadratic term of the action for gauge field becomes,
\bge
\label{GFaction}
  S_V+S_{GF}=\FR{1}{2}\int\di^4x\,\bigg[A^\mu(\eta_{\mu\nu}\pd^2-\pd_\mu\pd_\nu)A^\nu+\FR{1}{\zeta}A^\mu\pd_\mu \pd_\nu A^\nu\bigg],
\ede
in which the indices for $A^\mu$ is raised by flat metric $\eta^{\mu\nu}$. Then we are free to choose the gauge parameter $\zeta=1$ so that the equation of motion for $A_\mu$ takes the very simple form $\pd^2 A_\mu=0$, so does its solution,
\bge
  A_{\mu}(\tau,\mb x)=\int\FR{\di^3k}{(2\pi)^3}\FR{1}{2k}\sum_\lam\Big[e^{-\ii k\tau}\ep_\mu^\lam(\mb k)a_\lam(\mb k)+e^{\ii k\tau}\ep_\mu^{\lam*}(-\mb k)a^\dag_\lam(-\mb k)\Big]e^{\ii\mb k\cdot\mb x},
\ede
where $\ep_{\mu}^\lam(\mb k)$ is the polarization vector, $a_\lam(\mb k)$ is the corresponding annihilation operator, and $\lam$ labels $A_\mu$'s physical polarizations. It is also easy to work out the various tree-level in-in propagators for gauge field $A_\mu$,
\begin{align}
&\Big[G_{++}^{(V)}\Big]_{\mu\nu}(\mb k,\tau,\tau')=\FR{\ii\eta_{\mu\nu}}{2k}\Big[\theta(\tau-\tau')e^{-\ii k(\tau-\tau')}+\theta(\tau'-\tau)e^{-\ii k(\tau'-\tau)}\Big],\\
&\Big[G_{+-}^{(V)}\Big]_{\mu\nu}(\mb k,\tau,\tau')=\FR{\ii\eta_{\mu\nu}}{2k}e^{-\ii k(\tau'-\tau)},\\
&\Big[G_{-+}^{(V)}\Big]_{\mu\nu}(\mb k,\tau,\tau')=\FR{\ii\eta_{\mu\nu}}{2k}e^{-\ii k(\tau-\tau')},\label{PhoMPP}\\
&\Big[G_{--}^{(V)}\Big]_{\mu\nu}(\mb k,\tau,\tau')=\FR{\ii\eta_{\mu\nu}}{2k}\Big[\theta(\tau-\tau')e^{-\ii k(\tau'-\tau)}+\theta(\tau'-\tau)e^{-\ii k(\tau-\tau')}\Big],
\end{align}
in which we have substituted the polarization sum $\sum_\lam \ep_\mu^\lam(\mb k)\ep_\nu^{\lam*}(\mb p)\to\eta_{\mu\nu}$. This is not an equality on its own, but we can make such substitution as long as the photon couples to conserved sources.

Now let's consider briefly the general case when the spacetime dimension $d+1$ deviates from 4. Our calculation in the current paper actually does not need this result, but we still present them for completeness and for future reference. Then the gauge-fixed action (\ref{GFaction}) will be
\begin{align}
  S_V+S_{GF}=\FR{1}{2}\int\di^{d+1}x\,a^{d-3}\bigg[A^\mu(\eta_{\mu\nu}\pd^2-\pd_\mu\pd_\nu)A^\nu+\FR{1}{\zeta}A^\mu\pd_\mu \pd_\nu A^\nu-\FR{d-3}{\tau}A_\mu F^{0\mu}\bigg],
\end{align}
in which all indices are raised and lowered by flat metric. The additional prefactor $a^{d-3}$ and a new term proportional to $(d-3)$ shows the explicit breaking of Weyl invariance of gauge action when $d\neq3$. To find the equation of motion, we still choose the gauge $\zeta=1$, then the field equation for temporal component $A^0$ remains the same, while the equation for spatial component $A^i$ has new terms,
\begin{align}
&\pd^2A^0=0, &&\pd^2A^i+\FR{d-3}{\tau}(\pd_\tau A^i+\pd_i A^0)=0.
\end{align}
Finally we consider a massive photon in $(3+1)$d dS, with the following bare mass term in the action,
\bge
  S_V(M)=-\FR{1}{2}M^2\int\di^4x\,\sqrt{-g}g^{\mu\nu}A_\mu A_\nu.
\ede
Then in $\zeta=1$ gauge, the equation of motion becomes,
\bge
  \pd^2 A^\mu(x) - \FR{M^2}{H^2\tau^2}A^\mu(x)=0.~~~
  \To~~~
  A_{\mb k}^\mu{}''(\tau)+k^2 A_{\mb k}(\tau)+\FR{M^2}{H^2\tau^2}A^\mu_{\mb k}(\tau)=0.
\ede
So the solution is identical to a massive scalar field (\ref{MSM}) but with $d$ formally set to 1,
\bge
\label{MPM}
  A_{\mb k}(\tau)=-\ii\FR{\sqrt\pi}{2}e^{\ii \pi(\nu_A/2+1/4)}(-\tau)^{1/2}\mathrm{H}_{\nu_A}^{(1)}(-k\tau),
\ede
with $\nu_A\equiv \sqrt{1/4-(M/H)^2}$.

\section{One-Loop Correction to In-In Propagators}
\label{sec4}

In this section we are going to calculate 1-loop corrections to 2-point function of a real scalar, a Dirac spinor, and a photon field, respectively. For simplicity, we assume the following interactions for the real scalar field $\phi$ and the Dirac spinor $\psi$,
\bge
\label{Sint}
  S_{\text{int}}=\int\di^{d+1} x\,\sqrt{-g}\bigg[-\FR{\lam\mu^{3-d}}{24}\phi^4+y\mu^{(3-d)/2}\phi\ob{\psi}\psi\bigg],
\ede
where we have introduced a mass scale $\mu$ so that the scalar self-coupling $\lam$ and Yukawa coupling $y$ remain dimensionless. On the other hand, in order to couple the scalar field to the photon, we identify $\phi$ to be the real part of a complex scalar field $\Phi$, i.e., $\Phi=\frac{1}{\sqrt{2}}(\phi+\ii\pi)$, then the usual gauge interactions can be readily extracted from the kinetic terms,
\bge
\label{Sk}
  S_{\text{k}}=\int\di^{d+1}x\,\sqrt{-g}\bigg[g^{\mu\nu}(\D_\mu\Phi)^\dag(\D_\nu\Phi)+\FR{\ii}{2}\Big(\ob{\psi}\ga^\mu\mathscr{D}_\mu\psi-\ob\psi\overleftarrow{\mathscr{D}}_\mu\gamma^\mu\psi\Big)\bigg],
\ede
where $\D_\mu\Phi$ is the usual gauge covariant derivative, and $\mathscr{D}_\mu\psi$ is covariant derivative associated with both gauge transformation and local Lorentz transformation. In addition, if the gauge group is non-Abelian, we still have gauge self-interaction, but this is irrelevant to our study as it cannot generate any late-time divergence.

Before getting into the technical part of loop computations, we would like to comment here some general properties of loop correction that can be inferred from the result of last section. We have seen that among massless spin-$(0,\frac{1}{2},1)$ fields, the scalar is special in that the minimally coupled scalar field is not Weyl invariant, and thus its mode function in dS background deviates significantly from the flat space result. On the contrary, both the massless spin-1/2 and massless spin-1 fields have Weyl invariant kinetic term, which thus means that their mode function as well as the propagator in dS can be directly got from the flat space result by simple rescaling.

The above observation can be put in a more physical way, that is, the scalar field is more sensitive to the exponential expansion of the universe that spinor and vector field. As a result, its propagator will get stretched at large distances. In field theory we are familiar with the fact that whenever the propagator fails to die away fast enough at large distance, late-time divergence could appear, and in many situation, such late-time divergence can be resummed and contribute a nonzero mass term. A well-known example of this kind is the scalar QED in $(1+1)$d. In this case, the massless scalar propagator behaves like $\int \di^2k e^{\ii kx}/k^2\sim \log x$, and it does generate a nonzero mass for the photon. In the following we are going to show that spinor and photon loops don't contribute late-time divergences, and all such divergences in dS are from the massless scalar field running in the loop. However, they cannot always be interpreted as a signal of mass generation.

On the other hand, for processes involving fermions and photons only, there cannot be late-time divergence, and this fact can be most directly seen by counting the power of scalar factor $a(\tau)$. In fact, the generic late-time divergence arises due to the presence of factor $\sqrt{-g}=(H\tau)^{-4}$ in the interaction vertex. In this case the integral in the in-in calculation $\int\di\tau/\tau^4$ will generically be divergent, unless there are enough positive powers of $|\tau|$, or equivalently, enough negative powers of $a(\tau)$, to cancel the $\tau^{4}$ factor in the denominator. For instance, the scalar self-interaction, namely the first term of (\ref{Sint}) has nothing to cancel the prefactor $\sqrt{-g}$ and thus the corresponding $\tau$-integral is badly divergent as $\tau\to 0$.

On the contrary, the fermion interacts with photon through standard gauge coupling, which is derived from the last two terms of (\ref{Sk}) and reads $\sqrt{-g}e_m^\mu A_\mu\ob{\psi}\gamma^m\psi$. Now we have one factor of $a^{-1}$ coming from $e_m^\mu$ and additional 3 factors of $a^{-3}$ from fermion rescaling $\phi=(H\tau)^{3/2}\wt{\psi}$, so that the prefactor $\sqrt{-g}=a^4$ is precisely canceled, and therefore the fermion gauge interaction is free from late-time divergence. Similar counting applies for self-coupling of gauge boson when the gauge group is non-Abelian.

We would also like to make a technical remark on the structure of late-time divergence of 1-loop diagrams and to explain our strategy of evaluating them. All 1-loop diagrams in our calculation can be classified into 2 sets. In the first set, the loop is attached to 1 vertex only, like the diagrams in Fig.\,\ref{fig_phi4loop}. The other set consists of diagrams in which the loop is attached to 2 vertices, like the diagrams in Fig.\,\ref{fig_scalarfermionloop}. In the former case, the loop is simply given by $G(x,x)$. Since the Green function $G(x,y)$ depends only on the dS invariant distance, the loop factor $G(x,x)$ is dS invariant. On the other hand, it contains the usual ultraviolet (UV) divergences which is to be removed by corresponding local counterterm. Therefore, such loop factor cannot generate late-time divergences. In the following calculation, this loop factor can be regarded as a mass insertion, and the associated late-time divergence is actually from the improper perturbative expansion treating the mass term as an interaction, as has long been noticed \cite{Weinb05,Weinb06}. So the resummation of such divergences will naturally lead to an effective mass the the field being considered. Then it is easy to see that the coefficient of the late-time divergence is related to the UV divergences of the loop factor. Therefore, we use the standard dimensional regularization for such loop diagrams.

On the other hand, in the second set of diagrams, the loop factor is proportional to $G^2(x,y)$, and the distance between $x,y$ can in principle be arbitrary. In particular, when the loop modes become very soft at late times, they can generate additional late-time divergences which are more divergent than perturbative expansion of mass term, and these divergences have nothing to do with UV divergences. Therefore, we can actually get the leading late-time divergence without worrying about UV and renormalization. So for these diagrams, we can evaluate the leading late-time divergences by cut off the momentum integral at some particular scale without invoking dimensional regularization. A natural cut-off would be the external momentum of the 2-point function, and as we shall see below, the coefficient of leading late-time divergence is sensitive only to zero loop momentum limit, and thus is independent of the choice of this ``UV'' cut-off.

\subsection{Scalar Propagator}

\subsubsection{Scalar Loop}

\begin{figure}[tbph]
\centering
\includegraphics[width=0.6\textwidth]{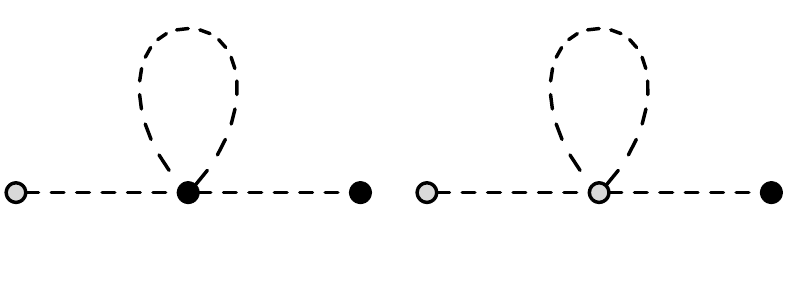}
\vspace{-10mm}
\emph{
\caption{1-loop correction to scalar 2-point function from scalar loop via $\phi^4$ interaction in Schwinger-Keldysh formalism.}
\label{fig_phi4loop}
}
\end{figure}

In the unbroken phase of electroweak symmetry in SM, the Higgs field only has quartic self-interaction, which can thus be well modeled by $\lam\phi^4$ theory of a real scalar. Now we calculate the 1-loop correction from this $\phi^4$ interaction to the two point function $\la\phi(\tau_1,\mb k_1)\phi(\tau_2,\mb k_2)\ra$, namely the propagator $G_{-+}$ defined in last section. In Schwinger-Keldysh double field formalism, there are two diagrams contributing to this process, shown in Fig.\,\ref{fig_phi4loop}.

In these diagrams, we use grey dots and black dots to represent $-$ and $+$ type vertices, respectively. An advantage of this diagrammatic representation is that we can easily find the complex conjugate of a diagram by turning black dots into grey and grey dots into black. When the two external vertices are taken to be at the same time, the sum of all possible diagrams is manifestly real. In this case, only internal vertices needs to be switched to get the complex conjugate of a single diagram. (However, additional care must be taken when applying this rule to diagrams containing fermions.)

The $\phi^4$ vertex of plus type with $d$ spatial dimensions is given by (\ref{Sint}), and the corresponding Feynman rule is,
\bge
  -\FR{\lam\mu^{3-d}}{24}\int\di^{d+1}x\sqrt{-g}\phi^4~~~~~\to~~~~~-\ii\sqrt{-g} \lam\mu^{3-d}=-\FR{\ii\lam\mu^{3-d}}{(H\tau)^{d+1}},
\ede
while the minus type has an additional minus sign. The introduction of a scale $\mu$ is to make the coupling $\lam$ dimensionless in $d=3-\ep$ spatial dimensions. It is straightforward then to calculate these two diagrams as,
\begin{align}
-\FR{1}{\ii}\de_{\text{1-loop}}\wh{G}_{-+}(\mb k,\tau_1,\tau_2)\big|_{\lam\phi^4}
  =-&\FR{1}{2}\cdot\ii\lam\mu^{3-d}\Big(\FR{1}{\ii}\Big)^3\int\FR{\di\tau'}{(H\tau')^{d+1}}\int\FR{\di^dq}{(2\pi)^d}\n\\
  \times\Big[&-\wh{G}_{-+}(\mb k,\tau_1,\tau')\wh{G}_{++}(\mb k,\tau',\tau_2)\wh{G}_{++}(\mb q,\tau',\tau')\n\\
  &+\wh{G}_{--}(\mb k,\tau_1,\tau')\wh{G}_{-+}(\mb k,\tau',\tau_2)\wh{G}_{--}(\mb q,\tau',\tau')\Big],
\end{align}
where the minus sign on the left hand side is from the  grey dot associated with one external vertex, the $1/\ii$ factor is from our normalization convention of Green function, and the factor $1/2$ on the right hand side comes from combinatorics of internal line. For simplicity but without loss of generality we can set $\tau_1=\tau_2\equiv \tau\to 0$. The loop momentum integral in above expression is regulated to be finite in the UV, but is still divergent in the IR. To deal with late-time divergence, we shall use massive mode functions $(\ref{MSM})$ for loop propagator, while for the external lines we still use massless mode functions (\ref{MLSM}) in $d=3$ spatial dimensions\footnote{One can keep external lines to be in general $d$ spatial dimensions too, but the additional terms arising in this case are canceled precisely by the corresponding contribution from counterterm if one also put counterterm diagrams consistently in $d$ spatial dimensions. This remark also applies to the $\tau'$-integral to be evaluated in the following.}. Then the above expression reduces to,
\begin{align}
\label{Sphi4loop}
&\ii\de_{\text{1-loop}}G_{-+}(\mb k,\tau_1,\tau_2)\big|_{\lam\phi^4}\n\\
  &=
  \FR{\ii\lam\mu^{3-d}H^4}{8k^6}\int_{-\infty}^{0}\FR{\di\tau'}{(H\tau')^{d+1}}\int\FR{\di^dq}{(2\pi)^d}u(\tau',\mb q)u^*(\tau',-\mb q)\Big[(1-\ii k\tau')^2e^{2\ii k\tau'}-(1+\ii k\tau')^2e^{-2\ii k\tau'}\Big]\n\\
  &=\FR{-\lam\mu^{3-d}H^4}{4k^6}\text{Im}\bigg[\int_{-\infty}^{0}\FR{\di\tau'}{(H\tau')^{d+1}}(1-\ii k\tau')^2e^{2\ii k\tau'}\int\FR{\di^dq}{(2\pi)^d}u(\tau',\mb q)u^*(\tau',-\mb q)\bigg],
\end{align}
where $u(\tau',\mb q)$ and $u^*(\tau',-\mb q)$ are given by (\ref{MSM}). The momentum integral now is convergent, and can be worked out explicitly as follows,
\begin{align}
\label{massivephi4loop}
&\mu^{3-d}\int\FR{\di^d q}{(2\pi)^d}u(\tau',\mb q)u^*(\tau',-\mb q)=\FR{\pi V_{d-1}}{4(2\pi)^{d}}H^{d-1}(-\tau)^{d}\int_0^\infty \di q\,q^{d-1}\Big|\mathrm{H}_{\nu}^{(1)}(-q\tau)\Big|^2\n\\
&=\FR{\mu^{3-d}H^{d-1}}{2^{d+1}\pi^{d/2-1}\Gamma(d/2)}\int_0^\infty\di z\,z^{d-1}\Big|\mathrm{H}_{\nu}^{(1)}(z)\Big|^2\n\\
=&\FR{\mu^{3-d}H^{d-1}\Gamma(\frac{d}{2}+\nu)}{4\pi^{d/2-1}\Gamma(d)\Gamma(1-\frac{d}{2}+\nu)}\bigg[2\csc\frac{\pi d}{2}\cot\frac{\pi(d-2\nu)}{2}+\sec\frac{\pi d}{2}\bigg],
\end{align}
where $V_{d-1}=2\pi^{d/2}/\Gamma(d/2)$ is the volume of $S^{d-1}$ of unit radius, in the second line we have defined $z\equiv -q\tau$, and we have made the assumption that the mass $m^2\leq\frac{9}{4}H^2$ so that $\nu$ is real. As we shall see, this is always true as long as the coupling is weak.

As expected, the loop momentum integral is $\tau'$-independent due to de Sitter invariance. Now we expand this expression in the $d=3-\ep\to 3$ and $m/H\to 0$ limit, which gives,
\bge
  \FR{H^2}{(2\pi)^2}\bigg[\FR{1}{\ep}+\FR{3H^2}{2m^2}+\log\FR{\mu}{H}+\order{\ep^0,m^0}\bigg].
\ede

On the other hand, we can also put $d=3$ in the $\tau'$-integral and  carry it out as,
\bge
  \int_{-\infty}^{\tau}\FR{\di\tau'}{\tau'^4}(1-\ii k\tau')^2e^{2\ii k\tau'}=-\FR{1}{3\tau^3}-\FR{k^2}{\tau}+\FR{2\ii k^3}{3}\bigg[\log(-2k\tau)-\FR{\ii\pi}{2}+\ga_E-\FR{7}{3}\bigg]+\order{\tau},
\ede
in which we have introduced a late-time cutoff $\tau$ in order to keep the result finite, and we also expand the result in terms of $\tau$ in the $\tau\to0$ limit. Then finally the sum of two diagrams from $\phi^4$ interaction is evaluated to,
\bge
  \FR{-\lam H^2}{6(2\pi)^2k^3}\bigg[\log(-2k\tau)+\ga_E-\FR{7}{3}\bigg]\bigg(\FR{1}{\ep}+\FR{3H^2}{2m^2}+\log\FR{\mu}{H}\bigg).
\ede

The $1/\ep$ pole corresponds to the UV divergence, which we can subtract by introducing counterterms. Let the counterterm in the Lagrangian be,
\bge
  -\FR{1}{2}\sqrt{-g}\de_m\phi^2(x),
\ede
then its contribution to the two point function can be easily found to be,
\begin{align}
\label{massct}
  &-\ii\de_m\Big(\FR{1}{\ii}\Big)^2\int\FR{\di\tau'}{(H\tau')^{d+1}}\Big[-\wh{G}_{-+}(\mb k;\tau_1,\tau')\wh{G}_{++}(\mb k;\tau',\tau_2)+\wh{G}_{--}(\mb k;\tau_1,\tau')\wh{G}_{-+}(\mb k;\tau',\tau_2)\Big]\n\\
  &\To\FR{\de_m}{3k^3}\bigg[\log(-2k\tau)+\ga_E-\FR{7}{3}\bigg],
\end{align}
where we have taken $d=3$ and $\tau_1=\tau_2=\tau\to 0$ in the second line. Then it is easy to see that we can choose $\de_m=\frac{\lam H^2}{2(2\pi)^2}\frac{1}{\ep}$ to eliminate the divergent terms coming from loop diagrams. This result is similar to the corresponding flat-space one, namely the 1-loop contribution from $\phi^4$ theory only contributes to mass renormalization but not to wave function renormalization. The difference is that the mass renormalization in flat space is proportional to the mass of the scalar field itself, and vanishes when the loop scalar is massless. But in our case, the counterterm does not vanish even for massless scalar due to the presence of another mass scale, namely the Hubble parameter of the background spacetime.

Then after subtracting the $\ep$-pole by counterterm, we reach the following final result,
\bge
\label{Sca1L}
-\ii\de_{\text{1-loop}}\wh{G}_{-+}(\mb k,\tau,\tau)\big|_{\lam\phi^4}=\FR{\lam H^2}{6(2\pi)^2k^3}\bigg[\log(-2k\tau)+\ga_E-\FR{7}{3}\bigg]\bigg(\FR{3H^2}{2m^2}+\log\FR{\mu}{H}\bigg).
\ede
The loop mass $m$ here acts as an effective IR cutoff. It is instructive to calculate the loop momentum integral using sharp cutoff regularization by introducing both UV cutoff $\Lambda_{\text{UV}}$ (which transfers to the renormalization scale $\mu$ after renormalization) and IR cutoff $\LIR$. Then the loop integral reads,
\begin{align}
\int_{\LIR}^{\Lambda_{\text{UV}}}\FR{\di^3q}{(2\pi)^3}\FR{H^2}{2q^3}(1+q^2\tau'^2)=\FR{H^2}{(2\pi)^2}\log\FR{\Lambda_{\text{UV}}}{\LIR}\To \FR{H^2}{(2\pi)^2}\log\FR{\mu a(\tau)}{\LIR},
\end{align}
where in the first equality we have thrown the $q^2\tau'^2$ term because 1) this term is regulated to a term quadratic in $\Lambda_{\text{UV}}$ and thus can be totally subtracted away by choosing counterterm properly and 2) this term is proportional to $\tau'^2$ and thus is more convergent as $\tau'\to 0$. Then in the second ``equality'' we have done the renormalization and transfer the UV cutoff $\Lambda_{\text{UV}}$ to the renormalization scale $\mu$. It is important to note that the cutoffs we introduced are for comoving momentum $k$ and thus are comoving scales too. On the other hand the renormalization scale $\mu$ should be a physical scale. Here by physical we don't mean physically observable, but only that it is defined with respect to local proper distance rather than comoving distance. Therefore, the correct ratio in the last expression should contain an additional scale factor $a(\tau)$. Now note that $a(\tau)=-(H\tau)^{-1}$, so we can compare the above expression with the result from dimensional regularization with finite mass $m$, and find the following correspondence,
\bge
\label{LIRm}
  -\log(-\LIR\tau)=\FR{3H^2}{2m^2}.
\ede
This relation between the comoving IR cutoff $\LIR$ and the scalar mass $m$ will be useful in the following, where we shall use it to convert $\LIR$ to $m$ without doing computation with massive loop fields.

\subsubsection{Fermion Loop}

\begin{figure}[tbph]
\centering
\includegraphics[width=0.6\textwidth]{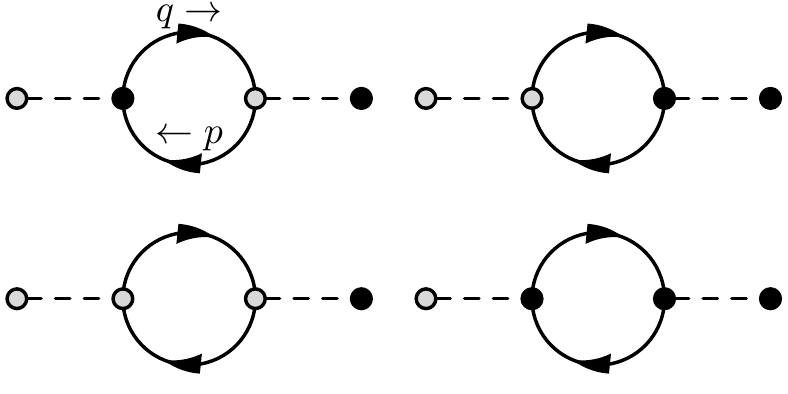}
\emph{
\caption{1-loop correction to scalar 2-point function from fermion loop via Yukawa interaction in Schwinger-Keldysh formalism.}
\label{fig_scalarfermionloop}
}
\end{figure}

In this subsection we consider the fermion loop contribution to the two point function of the scalar field $\phi$, via the Yukawa interaction:
\bge
  S_{Y}=y\mu^{(3-d)/2}\int\di^{d+1}x\,\sqrt{-g}\phi\ob\psi\psi~~~\to~~~ \FR{\ii y\mu^{(3-d)/2}}{(H\tau)^{d+1}}.
\ede
There are four diagrams correspond to the 1-loop correction to the scalar two-point function shown in Fig.\,\ref{fig_scalarfermionloop}, in which, again, grey dots represent minus type vertices and black dots represent plus type vertices. As commented at the beginning of this section, we can actually set $d=3$ as long as we are only concerned with leading late-time divergence. Then the sum of four diagrams reads,
{\allowdisplaybreaks
\begin{align}
&-\FR{1}{\ii}\de_{\text{1-loop}}\wh{G}_{-+}(\mb k,\tau_1,\tau_2)\big|_{\text{Yukawa}}= (\ii y)^2\Big(\FR{1}{\ii}\Big)^4\int\FR{\di\tau'}{(H\tau')^4}\FR{\di\tau''}{(H\tau'')^4}\int\FR{\di^3q}{(2\pi)^3}\n\\
 &~\times\bigg\{\wh{G}_{-+}(\mb k,\tau_1,\tau')\wh{G}_{-+}(\mb k,\tau'',\tau_2)(-)\tr\Big[\wh{G}_{-+}^{(F)}(\mb p,\tau'',\tau')\wh{G}_{+-}^{(F)}(\mb q,\tau',\tau'')\Big]\n\\
 &~~\;\,+\wh{G}_{--}(\mb k,\tau_1,\tau')\wh{G}_{++}(\mb k,\tau'',\tau_2)(-)\tr\Big[\wh{G}_{+-}^{(F)}(\mb p,\tau'',\tau')\wh{G}_{-+}^{(F)}(\mb q,\tau',\tau'')\Big]\n\\
 &~~\;\,-\wh{G}_{--}(\mb k,\tau_1,\tau')\wh{G}_{-+}(\mb k,\tau'',\tau_2)(-)\tr\Big[\wh{G}_{--}^{(F)}(\mb p,\tau'',\tau')\wh{G}_{--}^{(F)}(\mb q,\tau',\tau'')\Big]\n\\
 &~~\;\,-\wh{G}_{-+}(\mb k,\tau_1,\tau')\wh{G}_{++}(\mb k,\tau'',\tau_2)(-)\tr\Big[\wh{G}_{++}^{(F)}(\mb p,\tau'',\tau')\wh{G}_{++}^{(F)}(\mb q,\tau',\tau'')\Big]\bigg\},
\end{align}}%
together with the momentum conservation $\mb p=\mb q-\mb k$. As always we take $\tau_1=\tau_2\equiv \tau\to 0$ for clarity. Then the two upper diagrams of Fig.\;\ref{fig_scalarfermionloop} become complex conjugate of each other, but each of the two is actually manifestly real, so we can sum them up as,
\begin{align}
  &-\FR{ y^2H^2}{16k^6}\int_{-\infty}^{0}\FR{\di\tau'}{\tau'}\int_{-\infty}^0\FR{\di\tau''}{\tau''}\int\FR{\di^3 q}{(2\pi)^{3}}(1-\ii k\tau')(1+\ii k\tau'')\tr\bigg[\FR{\sla p\ob{\sla q}}{pq}\bigg]e^{\ii k(\tau'-\tau'')-\ii(p+q)(\tau''-\tau')}+\text{c.c.}\n\\
  &=-\FR{ y^2H^2}{2(2\pi)^3k^6}\int\di^3 q\Big(1+\FR{\mb q\cdot\mb p}{qp}\Big)\bigg|\int_{-\infty}^0\FR{\di\tau'}{\tau'} (1-\ii k\tau')e^{\ii (k+p+q)\tau'}\bigg|^2.
\end{align}
Here and in the following, a bar on 4-momentum variable $\ob{q}$ always means to flip the sign of the spatial components of unbarred variable $q$, and the $\gamma$-trace in above expression is given by,
\bge
  \tr\bigg[\FR{\sla p\ob{\sla q}}{pq}\bigg]=-4\FR{p\cdot\bar q}{pq}=4\bigg(1+\FR{\mb p\cdot\mb q}{pq}\bigg),
\ede
and it should be understood that $p=|\mb p|=|\mb q-\mb k|$ as a result of momentum conservation.

In the same way, the two lower diagrams of Fig.\;\ref{fig_scalarfermionloop} are also complex conjugates of each other, except that each of them is not necessarily real. So we also sum these two diagrams as,
\begin{align}
&\FR{2y^2H^2}{16k^6}\int_{-\infty}^{0}\FR{\di\tau'}{\tau'}\int_{-\infty}^{\tau'}\FR{\di\tau''}{\tau''}\int\FR{\di^3 q}{(2\pi)^3}(1+\ii k\tau')(1+\ii k\tau'')\tr\bigg[\FR{\ob{\sla p}{\sla q}}{pq}\bigg]e^{-\ii k(\tau'+\tau'')-\ii(q+p)(\tau''-\tau')}+\text{c.c.}\n\\
&=\FR{y^2H^2}{(2\pi)^3k^6}\int\di^3 q\Big(1+\FR{\mb q\cdot\mb p}{qp}\Big)\text{Re}\bigg[\int_{-\infty}^{0}\FR{\di\tau'}{\tau'}\int_{-\infty}^{\tau'}\FR{\di\tau''}{\tau''}(1+\ii k\tau')(1+\ii k\tau'')e^{-\ii k(\tau'+\tau'')-\ii(q+p)(\tau''-\tau')}\bigg]
\end{align}
Now we are ready to sum up the four diagrams together, which is,
\begin{align}
  -\FR{2y^2H^2}{2(2\pi)^3k^6}\int\di^{3}q\Big(1+\FR{\mb q\cdot\mb p}{qp}\Big)\mathcal{I}^{(F)}(q,k),
\end{align}
where the fermionic $\tau$-integral $\mathcal{I}^{(F)}(q,k)$ is given by,
\begin{align}
  \mathcal{I}^{(F)}(q,k)\equiv&~\bigg|\int_{-\infty}^\tau\FR{\di\tau'}{\tau'}(1-\ii k\tau')e^{\ii (k+q+p)\tau'}\bigg|^2\n\\
  &~-2\Re\int_{-\infty}^{\tau}\FR{\di\tau'}{\tau'}\int_{-\infty}^{\tau'}\FR{\di\tau''}{\tau''}(1+\ii k\tau')(1+\ii k\tau'')e^{-\ii k(\tau'+\tau'')-\ii(q+p)(\tau''-\tau')},
\end{align}
where we have introduced a late-time cutoff $\tau$ to the integral. Now that we are only concerned with late-time divergent part of this integral, we are free to expand the integrand in the $|k\tau'|,|p\tau'|,|q\tau'|\ll 1$ limit, and keep only the terms that would contribute to late-time divergences. On the other hand, due to this expansion, the integral will no longer be convergent as $\tau'\to -\infty$. Therefore we should introduce an early-time cutoff $\tau_e$ for the integral. A natural choice of $\tau_e$ is the time  when the $k$-mode is stretched outside the horizon, namely $\tau_e=-1/k$. Then we can approximate $\mathcal{I}^{(F)}(q,k)$ as,
\begin{align}
\mathcal{I}^{(F)}(q,k)=\bigg|\int_{-1/k}^\tau\FR{\di\tau'}{\tau'}\bigg|^2-2\text{Re}\int_{-1/k}^{\tau}\FR{\di\tau'}{\tau'}\int_{-1/k}^{\tau'}\FR{\di\tau''}{\tau''}+\order{\tau^0}=\order{\tau^0}.
\end{align}
Therefore $\mathcal{I^{(F)}}(q,k)$ is in fact convergent as the late-time cutoff $\tau\to 0$, so the fermion loop does not contribute to late-time divergence to scalar's 2-point function.

\subsubsection{Vector Loop}
\label{subsecScaVecLoop}

\begin{figure}[tbph]
\centering
\includegraphics[width=0.6\textwidth]{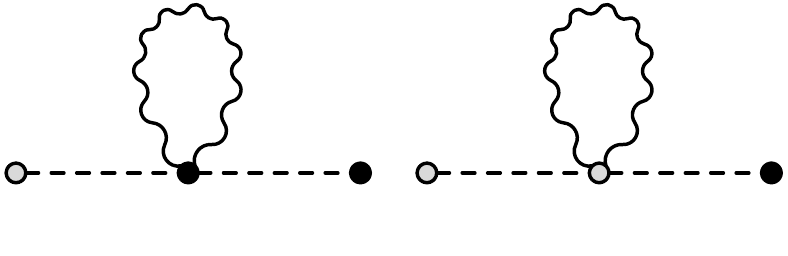}
\vspace{-6mm}
\emph{
\caption{1-loop correction to scalar 2-point function from photon loop via 4-point gauge interaction in Schwinger-Keldysh formalism.}
\label{fig_scalarphotonloop1}
}
\end{figure}

Now we identify the scalar field $\phi$ considered above in this section as the real part of a complex scalar field $\Phi=\frac{1}{\sqrt 2}(\phi+\ii\pi)$, in order to couple it to a $U(1)$ gauge field $A_\mu$. Then there are two types of vector 1-loop diagrams, corresponding to the following two types of terms expanded from the action (\ref{Sk}).
\bge
  S_\text{k}\supset-\int\di^4x\,\sqrt{-g}\bigg[\FR{1}{2}e^2g^{\mu\nu}A_\mu A_\nu \phi^2+e g^{\mu\nu}A_\mu(\phi\pd_\nu\pi-\pi\pd_\nu\phi)\bigg].
\ede
The first type of diagrams are shown in Fig.\,\ref{fig_scalarphotonloop1}, and can be readily calculated as,
\begin{align}
  -2\ii e^2\Big(\FR{1}{\ii}\Big)^3\int\FR{\di\tau'}{(H\tau')^2}\int\FR{\di^3q}{(2\pi)^3}\bigg\{&-\wh{G}_{-+}(\mb k,\tau_1,\tau')\wh{G}_{++}(\mb k,\tau',\tau_2)\Big[\wh{G}_{++}^{V}\Big]{}_\mu^\mu(q,\tau',\tau')\n\\
  &+\wh{G}_{--}(\mb k,\tau_1,\tau')\wh{G}_{-+}(\mb k,\tau',\tau_2)\Big[\wh{G}_{--}^{V}\Big]{}_\mu^\mu(q,\tau',\tau')\bigg\}
\end{align}
It can be immediately recognized that the contribution from these two diagrams is zero, since they contains the momentum integral $\int\di^3q/q$, which is regulated to zero under dimensional regularization.

\begin{figure}[tbph]
\centering
\includegraphics[width=0.6\textwidth]{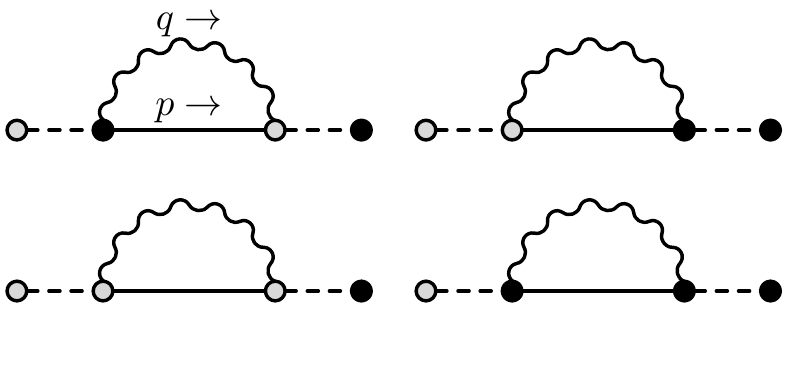}
\emph{
\caption{1-loop correction to scalar 2-point function from photon loop via 3-point gauge interaction in Schwinger-Keldysh formalism.}
\label{fig_scalarphotonloop2}
}
\end{figure}

Then we consider the second type of diagrams as shown in Fig.\,\ref{fig_scalarphotonloop2}, in which straight solid lines represent $\pi$ field, which has the same propagator with $\phi$. To compute these four diagrams, we need to treat temporal and spatial components of loop photon separately. Let's firstly look at the contributions from the temporal component. Then the four diagrams can be written as,
\begin{align}
&(-\ii e)^2\Big(\FR{1}{\ii}\Big)^4\int\FR{\di\tau'}{(H\tau')^2}\FR{\di\tau''}{(H\tau'')^2}\int\FR{\di^3q}{(2\pi)^3}\n\\
&\times\bigg\{\Big[\wh{G}_{+-}^{(V)}\Big]_{00}(\mb q,\tau',\tau'')\wh{G}_{-+}(\mb k,\tau_1,\tau')(\pd_{\tau'}-\overleftarrow{\pd}_{\tau'})\wh{G}_{+-}(\mb p,\tau',\tau'')(\overleftarrow{\pd}_{\tau''}-\pd_{\tau''})\wh{G}_{-+}(\mb k,\tau'',\tau_2)\n\\
&~\;\,+ \Big[\wh{G}_{-+}^{(V)}\Big]_{00}(\mb q,\tau',\tau'')\wh{G}_{--}(\mb k,\tau_1,\tau')(\pd_{\tau'}-\overleftarrow{\pd}_{\tau'})\wh{G}_{-+}(\mb p,\tau',\tau'')(\overleftarrow{\pd}_{\tau''}-\pd_{\tau''})\wh{G}_{++}(\mb k,\tau'',\tau_2)\n\\
&~\;\,- \Big[\wh{G}_{--}^{(V)}\Big]_{00}(\mb q,\tau',\tau'')\wh{G}_{--}(\mb k,\tau_1,\tau')(\pd_{\tau'}-\overleftarrow{\pd}_{\tau'})\wh{G}_{--}(\mb p,\tau',\tau'')(\overleftarrow{\pd}_{\tau''}-\pd_{\tau''})\wh{G}_{-+}(\mb k,\tau'',\tau_2)\n\\
&~\;\,- \Big[\wh{G}_{++}^{(V)}\Big]_{00}(\mb q,\tau',\tau'')\wh{G}_{-+}(\mb k,\tau_1,\tau')(\pd_{\tau'}-\overleftarrow{\pd}_{\tau'})\wh{G}_{++}(\mb p,\tau',\tau'')(\overleftarrow{\pd}_{\tau''}-\pd_{\tau''})\wh{G}_{++}(\mb k,\tau'',\tau_2)\bigg\},
\end{align}
in which the partial derivatives act only on adjacent factors. Without evaluation it is clear that this expression is free from late-time divergence, because each $\tau$-derivative of scalar mode is linear in $\tau$ variable, and thereby cancel one $\tau$ factor in the denominator. So this piece cannot generate mass for the scalar, and we then go on with the spatial component of loop photon, which gives the following expression,
\begin{align}
&(-\ii e)^2\Big(\FR{1}{\ii}\Big)^4\int\FR{\di\tau'}{(H\tau')^2}\FR{\di\tau''}{(H\tau'')^2}\int\FR{\di^3p}{(2\pi)^3}(-q_iq_j)\n\\
&\times\bigg\{\Big[\wh{G}_{+-}^{(V)}\Big]_{ij}(\mb q,\tau',\tau'')\wh{G}_{-+}(\mb k,\tau_1,\tau')\wh{G}_{+-}(\mb p,\tau',\tau'')\wh{G}_{-+}(\mb k,\tau'',\tau_2)\n\\
&~\;\,+\Big[\wh{G}_{-+}^{(V)}\Big]_{ij}(\mb q,\tau',\tau'')\wh{G}_{--}(\mb k,\tau_1,\tau')\wh{G}_{-+}(\mb p,\tau',\tau'')\wh{G}_{++}(\mb k,\tau'',\tau_2)\n\\
&~\;\,-\Big[\wh{G}_{--}^{(V)}\Big]_{ij}(\mb q,\tau',\tau'')\wh{G}_{--}(\mb k,\tau_1,\tau')\wh{G}_{--}(\mb p,\tau',\tau'')\wh{G}_{-+}(\mb k,\tau'',\tau_2)\n\\
&~\;\,-\Big[\wh{G}_{++}^{(V)}\Big]_{ij}(\mb q,\tau',\tau'')\wh{G}_{-+}(\mb k,\tau_1,\tau')\wh{G}_{++}(\mb p,\tau',\tau'')\wh{G}_{++}(\mb k,\tau'',\tau_2)\bigg\}.
\end{align}
We still evaluate this expression in the limit $\tau_1=\tau_2\equiv \tau\to 0$. Then the two diagrams in the first line sum to,
\bge
  -\FR{e^2H^2}{8k^6}\int\FR{\di^3p}{(2\pi)^3}\FR{-q}{p^3}\bigg|\int_{-\infty}^0\FR{\di\tau'}{\tau'^2}(1-\ii k\tau')(1-\ii p\tau')e^{\ii(q+k+p)\tau'}\bigg|^2,
\ede
and the two diagrams in the second line sum to,
\begin{align}
  &\FR{e^2H^2}{4k^6}\int\FR{\di^3p}{(2\pi)^3}\FR{-q}{p^3}\n\\
  &\times\text{Re}\bigg[\int_{-\infty}^0\FR{\di\tau'}{\tau'^2}\int_{-\infty}^{\tau'}\FR{\di\tau''}{\tau''^2}(1+\ii k\tau')(1-\ii p\tau')(1+\ii k\tau'')(1+\ii p\tau'')e^{\ii(q+p)(\tau'-\tau'')-\ii k(\tau'+\tau'')}\bigg],
\end{align}
Then we need to evaluate the $\tau'$-integral,
\begin{align}
  &\mathcal{I}^{(V)}(p,k;\tau)
  =\bigg|\int_{-\infty}^\tau\FR{\di\tau'}{\tau'^2}(1-\ii k\tau')(1-\ii p\tau')e^{\ii(q+k+p)\tau'}\bigg|^2\n\\
  &-2\text{Re}\bigg[\int_{-\infty}^\tau\FR{\di\tau'}{\tau'^2}\int_{-\infty}^{\tau'}\FR{\di\tau''}{\tau''^2}(1+\ii k\tau')(1-\ii p\tau')(1+\ii k\tau'')(1+\ii p\tau'')e^{\ii(q+p)(\tau'-\tau'')-\ii k(\tau'+\tau'')}\bigg],
\end{align}
where we introduce the late-time cutoff $\tau$ as before. As in the case of fermion loop considered before, we can again expand the integrand to extract the IR divergent part, as follows,
\begin{align}
  \mathcal{I}^{(V)}(p,k;\tau)=&~\bigg|\int_{-1/k}^\tau\FR{\di\tau'}{\tau'^2}(1+\ii q\tau')\bigg|^2-2\text{Re}\bigg[\int_{-1/k}^\tau\FR{\di\tau'}{\tau'^2}\int_{-1/k}^{\tau'}\FR{\di\tau''}{\tau''^2}(1+\ii q\tau')(1-\ii q\tau'')\bigg]+\order{\tau^0}\n\\
  =&~\order{\tau^0}.
\end{align}
Therefore, similar to fermion loop, the 1-loop correction from photon to scalar's 2-point function is also free from late-time divergence.

\subsection{Fermion Propagator}

\begin{figure}[tbph]
\centering
\includegraphics[width=0.6\textwidth]{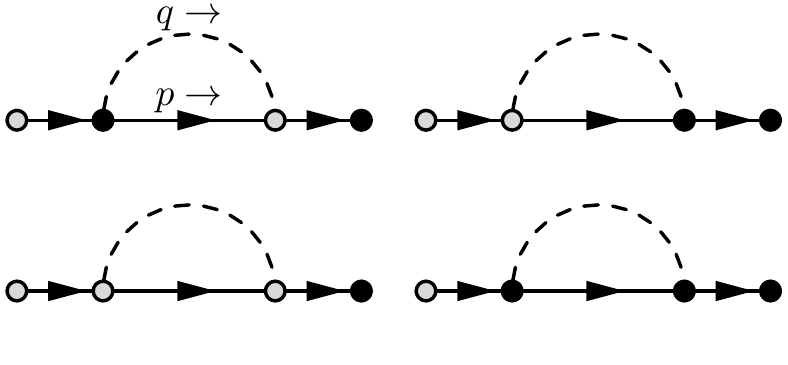}
\emph{
\caption{1-loop correction to Dirac fermion's 2-point function from scalar loop via Yukawa interaction in Schwinger-Keldysh formalism.}
\label{fig_fermionscalarloop}
}
\end{figure}

Now we consider the fermion 2-point function. There are two types of contribution at 1-loop order, with scalar loop via Yukawa interaction, and photon loop via gauge interaction, respectively. Without explicit calculation, we can already know that photon loop cannot generate mass to the fermion, as explained at the beginning of this section. So in current subsection, we only need to consider the scalar loop, as shown in Fig.\,\ref{fig_fermionscalarloop}.

These four diagrams can be collectively expressed as,
\begin{align}
&-\FR{1}{\ii}\de_{\text{1-loop}}\wh{G}^{(F)}_{-+}(\mb k,\tau_1,\tau_2) =(\ii y)^2\Big(\FR{1}{\ii}\Big)^4\int\FR{\di\tau'}{(H\tau')^4}\FR{\di\tau''}{(H\tau'')^4}\FR{\di^3p}{(2\pi)^3}\n\\
 &~\times\bigg\{\wh{G}_{+-}(\mb q,\tau',\tau'')\wh{G}_{-+}^{(F)}(\mb k,\tau'',\tau_2)\wh{G}_{+-}^{(F)}(\mb p,\tau',\tau'')\wh{G}_{-+}^{(F)}(\mb k,\tau_1,\tau')\n\\
 &~\,\;\;+\wh{G}_{-+}(\mb q,\tau',\tau'')\wh{G}_{++}^{(F)}(\mb k,\tau'',\tau_2)\wh{G}_{-+}^{(F)}(\mb p,\tau',\tau'')\wh{G}_{--}^{(F)}(\mb k,\tau_1,\tau')\n\\
 &~\,\;\;-\wh{G}_{--}(\mb q,\tau',\tau'')\wh{G}_{-+}^{(F)}(\mb k,\tau'',\tau_2)\wh{G}_{--}^{(F)}(\mb p,\tau',\tau'')\wh{G}_{--}^{(F)}(\mb k,\tau_1,\tau')\n\\
 &~\,\;\;-\wh{G}_{++}(\mb q,\tau',\tau'')\wh{G}_{++}^{(F)}(\mb k,\tau'',\tau_2)\wh{G}_{++}^{(F)}(\mb p,\tau',\tau'')\wh{G}_{-+}^{(F)}(\mb k,\tau_1,\tau')\bigg\}.
\end{align}
This is our first example of loop correction to fermions, so we show more details. Firstly the upper-left diagram of Fig.\;\ref{fig_fermionscalarloop} reads,
\begin{align}
&(\ii y)^2 \int\FR{\di \tau'}{(H\tau')^4}\FR{\di \tau''}{(H\tau'')^4}\int\FR{\di^3 p}{(2\pi)^3}\FR{H^2}{2q^3}(1+\ii q\tau'')(1-\ii q\tau')e^{-\ii q(\tau''-\tau')}\n\\
&\times\FR{H^9}{8k^2p}\big(\tau_1\tau_2\tau'^2\tau''^2\big)^{3/2}\sla k\ob{\sla p}\sla k e^{-\ii k(\tau''-\tau_2)-\ii p(\tau''-\tau')-\ii k(\tau_1-\tau')}\n\\
\To &\FR{y^2|H\tau|^3}{8k^2}\sla k\int\FR{\di^3 p}{(2\pi)^3}\FR{k\cdot\ob p}{pq^3}\bigg|\int\FR{\di\tau'}{\tau'}(1-\ii q\tau')e^{+\ii (q+p+k)\tau'}\bigg|^2,
\end{align}
where we used $\sla k\ob{\sla p}\sla k=-\sla k\sla k\ob{\sla p}-2k\cdot\ob{p}\sla k=-2k\cdot\ob{p}\sla k$, and we have taken $\tau_1=\tau_2=\tau$.  Then the upper-right diagram of Fig.\;\ref{fig_fermionscalarloop} is,
\bge
-\FR{y^2|H\tau|^3}{8k^2}\ob{\sla k}\int\FR{\di^3 p}{(2\pi)^3}\FR{\ob k\cdot p}{pq^3}\bigg|\int\FR{\di\tau'}{\tau'}(1-\ii q\tau')e^{+\ii (q+p+k)\tau'}\bigg|^2.
\ede
Sum the two diagrams together, we get,
\bge
-\FR{y^2|H\tau|^3}{4k^2}\mb k\cdot\bm{\ga}\int\FR{\di^3 p}{(2\pi)^3}\FR{k\cdot\ob p}{pq^3}\bigg|\int\FR{\di\tau'}{\tau'}(1-\ii q\tau')e^{+\ii (q+p+k)\tau'}\bigg|^2.
\ede

Now so long as we are concerned with the late-time divergence, we can expand the integrand of $\tau'$-integral as,
\bge
  (1-\ii q\tau')e^{\ii(q+p+k)\tau'}=1+\ii(p+k)\tau'+\order{\tau'^2},
\ede
so that all neglected $\order{\tau'^2}$ contribute no divergent terms as the late-time cutoff $\tau$ is sent to zero. On the other hand, we should note that the condition $|k\tau'|,|p\tau'|,|q\tau'|\ll 1$ that we have assumed in doing expansion is not valid for all $\tau'\ll \tau$. In particular, the integral is divergent as $\tau'\to-\infty$ when we use the expanded integrand. Therefore we should cut off the time integral at some early moment $\tau_e$. For our purpose it is natural to choose $\tau_e$ to be the moment of horizon exit of mode $k$. Meanwhile, we also note that the loop momenta $p,q\equiv |\mb k-\mb p|$ would satisfy $|p\tau'|,|q\tau'|\ll 1$ for all $\tau'\in(\tau_e,\tau)$ only when $p,q<k$. Correspondingly, we should also impose a UV cutoff for the momentum integral at the scale of $k$. The physical interpretation of our prescription for the range of integration is that we are considering the quantum correction from the modes outside the horizon at late times. At the same time we also need to introduce the IR cutoff $\LIR$ to eliminate the late-time divergence\footnote{When $p,q>k$ and $\tau'$ is very small, the condition $|p\tau|,|q\tau|\ll 1$ can still be satisfied, and the integral of this parameter region may also lead to some late-time divergence. However, as will be clear in the following, the late-time divergence from this region is only sub-leading, compared with the contribution from $p,q\ll k$, because in the latter case there will be additional $\log(\LIR)$ term which will give another factor of $\log(-\tau)$. Since we are only concerned with leading divergence in this paper, we are allowed to ignore this parameter region in the loop integral.}.

Now with above points clarified now we can carry out the time integral to be,
\bge
\bigg|\int_{-1/k}^{\tau}\FR{\di\tau'}{\tau'}\Big(1+\ii(p+k)\tau'\Big)\bigg|^2\simeq \log^2(-k\tau)+\order{\tau^0}.
\ede
Then we can finish the momentum integral,
\begin{align}
&\int\FR{\di^3p}{(2\pi)^3}\FR{k\cdot\ob{p}}{pq^3}=\FR{1}{(2\pi)^2}\int_{-1}^{1}\di\cos\theta\int_{\LIR}^{k-\LIR}\di p\, p^2\FR{-kp(1+\cos\theta)}{p(k^2+p^2-2kp\cos\theta)^{3/2}}\n\\
&=\FR{1}{(2\pi)^2}\Big[3(k-2\LIR)-4k\,\mathrm{arctanh}\Big(1-\FR{2\LIR}{k}\Big)\Big]=-\FR{2k}{(2\pi)^2}\log\FR{k}{\LIR}+\order{\LIR^0}.
\end{align}
Here in the first line we have set the upper limit of the integral to $k-\LIR$ to eliminate the late-time divergence from $q\equiv |\mb k-\mb p|\to 0$, and in the last equality we have done the expansion around $\LIR=0$. Now using the relation (\ref{LIRm}) we got earlier, we reach the following result,
\begin{align}
\label{FerYuLoop12}
\FR{y^2|H\tau|^3}{(2\pi)^2k}\mb k\cdot\bm{\ga}\bigg[\log(-k\tau)+\FR{3H^2}{2m^2}\bigg]\log^2(-k\tau).
\end{align}

Now let's consider the lower-left diagram of Fig.\;\ref{fig_fermionscalarloop},
\begin{align}
&(\ii y)^2\int\FR{\di \tau'}{(H\tau')^4}\int^{\tau'}\FR{\di \tau''}{(H\tau'')^4}\int\FR{\di^3p}{(2\pi)^3}\FR{H^2}{2q^3}(1+\ii q\tau'')(1-\ii q\tau')e^{-\ii q(\tau''-\tau')}\n\\
&\times\FR{H^9}{8k^2p}\big(\tau_1\tau_2\tau'^2\tau''^2\big)^{3/2}\sla k\ob{\sla p}\ob{\sla k} e^{-\ii k(\tau''-\tau_2)-\ii p(\tau''-\tau')-\ii k(\tau'-\tau_1)}\n\\
=&-(\ii y)^2\int\FR{\di \tau''}{(H\tau'')^4}\int^{\tau''}\FR{\di \tau'}{(H\tau')^4}\int\FR{\di^3p}{(2\pi)^3}\FR{H^2}{2q^3}(1+\ii q\tau')(1-\ii q\tau'')e^{-\ii q(\tau'-\tau'')}\n\\
&\times\FR{H^9}{8k^2p}\big(\tau_1\tau_2\tau'^2\tau''^2\big)^{3/2}\sla k {\sla p}\ob{\sla k} e^{-\ii k(\tau''-\tau_2)-\ii p(\tau'-\tau'')-\ii k(\tau'-\tau_1)}\n\\
\To&~\FR{-y^2|H\tau|^3}{16k^2}\int\FR{\di^3p}{(2\pi)^3}\FR{1}{q^3p}\big[\sla k\ob{\sla p}\ob{\sla k}-\sla k {\sla p}\ob{\sla k}\big]\n\\
&~\times\int\FR{\di \tau'}{\tau'}\int^{\tau'}\FR{\di \tau''}{\tau''}(1+\ii q\tau'')(1-\ii q\tau')e^{-\ii (q+p)(\tau''-\tau')-\ii k(\tau''+\tau')}
\end{align}
Similarly, the lower-right diagram of Fig.\;\ref{fig_fermionscalarloop} gives,
\begin{align}
&\FR{-y^2|H\tau|^3}{16k^2}\int\FR{\di^3p}{(2\pi)^3}\FR{1}{q^3p}\big[\ob{\sla k}{\sla p}{\sla k}-\ob{\sla k}\ob{\sla p}{\sla k}\big]\n\\
&~\times\int\FR{\di \tau'}{\tau'}\int^{\tau'}\FR{\di \tau''}{\tau''}(1+\ii q\tau')(1-\ii q\tau'')e^{-\ii (q+p)(\tau'-\tau'')+\ii k(\tau''+\tau')}
\end{align}
But one can readily show that the gamma matrices products in above expressions vanish. For instance, in the lower-left diagram of Fig.\;\ref{fig_fermionscalarloop} we have,
\begin{align}
\label{gammafactor1}
\sla k\ob{\sla p}\ob{\sla k}-\sla k\sla p\ob{\sla k}
=\sla k(-\ob{\sla k}\ob{\sla p}-2\ob k\cdot\ob p)-\sla k(-\ob{\sla k}\sla p-2\ob{k}\cdot p)=2\sla k\ob{\sla k}\mb p\cdot\bm{\ga}-4\mb k\cdot\mb p\sla k.
\end{align}
Then, as long as $p$ is a variable to be integrated over, we are allowed to make the substitution $\mb p\cdot\bm{\ga}\To (\mb k\cdot \mb p)(\mb k\cdot\bm{\ga})/k^2$. To simplify the expression further, we can assume without loss of generality that $\mb k=(0,0,k)$ is in $x^3$ direction, so we have $\sla k\ob{\sla k}=k^2(2-[\ga^3,\ga^0])$. Then it is clear that (\ref{gammafactor1}) vanishes, and so does the factor in the lower-right diagram of Fig.\;\ref{fig_fermionscalarloop}. So we conclude that the 1-loop correction for massless scalar field to the Dirac fermion in the late-time limit is given by (\ref{FerYuLoop12}), that is,
\bge
\label{Fer1L}
  -\ii\de_{\text{1-loop}}\wh{G}^{(F)}_{-+}(\mb k,\tau,\tau)=-\FR{y^2|H\tau|^{3}}{(2\pi)^2k}\mb k\cdot\bm{\ga}\bigg[\log(-k\tau)+\FR{3H^2}{2m^2}\bigg]\log^2(-k\tau).
\ede

\subsection{Photon Propagator}

Just like photon loop cannot contribute late-time divergence to fermion's propagator, fermion loop won't generate late-time divergence for photon propagator either, as a consequence of simple power counting of conformal time variable. Therefore we only need to consider scalar loops, which come in two categories, corresponding to two interactions expanded from (\ref{Sk}),
\bge
  S_\text{k}\supset-\int\di^4x\,\sqrt{-g}\bigg[\FR{1}{2}e^2g^{\mu\nu}A_\mu A_\nu( \phi^2 +\pi^2)+e g^{\mu\nu}A_\mu(\phi\pd_\nu\pi-\pi\pd_\nu\phi)\bigg].
\ede

\begin{figure}[tbph]
\centering
\includegraphics[width=0.6\textwidth]{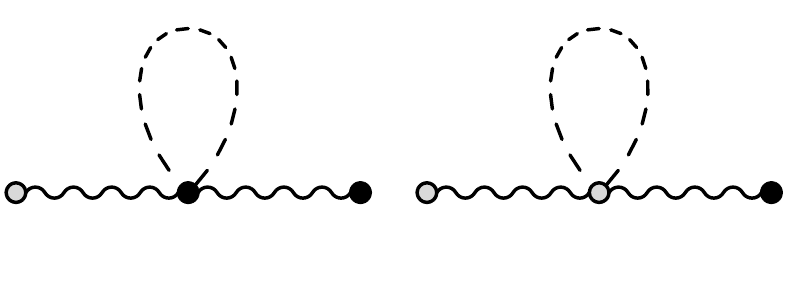}
\vspace{-6mm}
\emph{
\caption{1-loop correction to photon's 2-point function from scalar loop via 4-point gauge interaction in Schwinger-Keldysh formalism.}
\label{fig_photonscalarloop1}
}
\end{figure}

Let's consider the 4-point interaction first, as shown in Fig.\,\ref{fig_photonscalarloop1}, where only $\phi$-loops are shown explicitly, but it should be understood that there are still a pair of $\pi$-loop diagrams, which give identical results as $\phi$-loops. This gives an extra factor of 2, but we still have a combinatorial factor $1/2$, so taking all of them into account, we have,
\begin{align}
\label{Pho1L4}
-\FR{1}{\ii}\de_{\text{1-loop}}\Big[\wh{G}^{(V)}_{-+}\Big]_{\mu\nu}&(\mb k,\tau,\tau)\Big|_{\text{scalar,4p}}
=-2\ii e^2\mu^{3-d}\Big(\FR{1}{\ii}\Big)^3\int\FR{\di\tau'}{(-H\tau')^{d-1}}\int\FR{\di^dq}{(2\pi)^d}\n\\
&\times\bigg\{-\Big[\wh{G}_{-+}^{(V)}\Big]_{\mu\rh}(\mb k,\tau_1,\tau')\Big[\wh{G}_{++}^{(V)}\Big]^{\rh}_{\nu}(\mb k,\tau',\tau_2)\wh{G}_{++}(\mb q,\tau',\tau')\n\\
&~~~~~~+\Big[\wh{G}_{--}^{(V)}\Big]_{\mu\rh}(\mb k,\tau_1,\tau')\Big[\wh{G}_{-+}^{(V)}\Big]^{\rh}_{\nu}(\mb k,\tau',\tau_2)\wh{G}_{--}(\mb q,\tau',\tau')\bigg\}\n\\
&=\FR{e^2}{2k^2}\mu^{\ep}\eta_{\mu\nu}\,\cdot2\text{Im}\,\bigg[\int\FR{\di\tau'}{(-H\tau')^{2-\ep}}e^{-2\ii k\tau'}\int\FR{\di^{3-\ep}q}{(2\pi)^{3-\ep}}u(\tau',\mb k)u^*(\tau',-\mb k)\bigg]\n\\
&\To-\FR{e^2}{(2\pi)^2k}\cdot\FR{3H^2}{m^2}\log(-k\tau)\eta_{\mu\nu},
\end{align}
where the massive mode functions $u(\tau',\mb k)$ and $u^*(\tau',-\mb k)$ are given by (\ref{MSM}).
We see that the above scalar loops induce late-time divergence for photon's propagator, in a similar way as the loop correction in $\phi^4$ theory. In above expression we have neglected terms which contribute to additional $1/\ep$ but are fully subtracted by counterterms and thus leave nothing in finite terms. These terms are from the expansion of scalar propagator, as discussed in the calculation of $\phi^4$ loop.

\begin{figure}[tbph]
\centering
\includegraphics[width=0.6\textwidth]{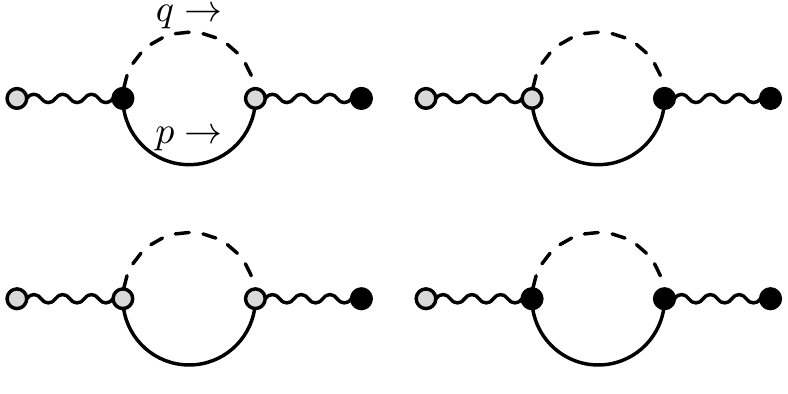}
\emph{
\caption{1-loop correction to photon's 2-point function from scalar loop via 3-point gauge interaction in Schwinger-Keldysh formalism.}
\label{fig_photonscalarloop2}
}
\end{figure}

Next we consider the 3-point interactions as shown in Fig.\,\ref{fig_photonscalarloop2}. Once again, it is important to distinguish the $A_0$-propagator and $A_i$-propagator, because the time and spatial derivative of loop scalar modes have very different forms. For $A_0$-propagator to which the time derivative mode $\dot\phi,\dot\pi\sim\tau$ contribute, the diagrams cannot generate late-time divergences, as can be seen by simple counting (See Section \ref{subsecScaVecLoop}). So we only need to consider the spatial components of the propagator,
The leading divergence of the four diagrams can be directly summed as,
\begin{align}
-\FR{1}{\ii}\de_{\text{1-loop}}\Big[\wh{G}^{(V)}_{-+}\Big]_{ij}&(\mb k,\tau,\tau)\Big|_{\text{scalar,3p}}=-\FR{e^2}{4k^2}\int\FR{\di^{3}p}{(2\pi)^3}\FR{1}{2q^3}\FR{1}{2p^3}(p_i-q_i)(p_j-q_j)\times2\mathcal{I}_V^{(S)}(k,p;\tau),
\end{align}
where the $\tau$-integral is given by
\begin{align}
\mathcal{I}_V^{(S)}(k,p;\tau)=&~\bigg|\int^\tau\FR{\di\tau'}{(-\tau')^{2}}(1+\ii q\tau')(1+\ii p\tau')e^{-\ii(q+p+k)\tau'}\bigg|^2\n\\
&-2\text{Re}\bigg[\int^\tau\FR{\di\tau'}{(-\tau')^{2}}\int^{\tau'}\FR{\di\tau''}{(-\tau'')^{2}}(1+\ii q\tau')(1+\ii p\tau')\n\\
&~~~~~\times(1-\ii q\tau'')(1-\ii p\tau'')e^{-\ii(q+p)(\tau'-\tau'')+\ii k(\tau'+\tau'')}\bigg].
\end{align}
Then we can carry out the integral $\mathcal{I}_V^{(S)}$ with the late-time cutoff $\tau\to 0$. The result is divergent as $\log\tau$. To identify these divergences, we can expand the integrand of $\mathcal{I}_V^{(S)}(k,p;\tau)$ as follows,
\begin{align}
  &(1+\ii q\tau')(1+\ii p\tau')e^{-\ii(q+p+k)\tau'}=1-\ii k\tau'+\order{\tau'^2},
\end{align}
then the Integral $\mathcal{I}_V^{(S)}$ reads,
\begin{align}
\mathcal{I}_V^{(S)}(k,p;\tau)=2k^2\log^2(-k\tau)+(p^2+q^2-k^2)\log(-k\tau)+\order{\tau^0}.
\end{align}
It should be noted that the condition $|k\tau'|,|p\tau'|,|q\tau'|\ll 1$ is not satisfied for any $\tau'<\tau$ and any $k,p,q=|\mb k-\mb p|$. In particular, the $\tau'$-integral is divergent as $\tau'\to -\infty$ when we do expansion as above. Therefore we have to cut off this integral at some early time $\tau_e$. Then the above result corresponds to choosing this early-time cutoff $\tau_e$ to be the time of horizon exit for mode $k$. Meanwhile, the condition $|p\tau|\ll 1$ and $|q\tau|\ll 1$ can be satisfied only when $p,q<k$. So in the momentum integral, we also need to introduce a UV cutoff for the loop momentum around $k$.

Now we insert this result back into the momentum integral, and define a new integration variable $\mb p'=\mb p-\mb q$, so that $\mb p=\frac{1}{2}(\mb k+\mb p')$ and $\mb q=\frac{1}{2}(\mb k-\mb p')$. In this way, the momentum integral can be rewritten as,
\begin{align}
  &\int\FR{\di^{3}p'}{8(2\pi)^3}\FR{1}{\frac{1}{4}|\mb k-\mb p'|^3}\FR{1}{\frac{1}{4}|\mb k+\mb p'|^3}p'_i p'_j\mathcal{I}_V^{(S)}(k,p;\tau)\n\\
  =&\int\FR{\di^{3}p'}{(2\pi)^3}\FR{\frac{1}{3}\eta_{ij}p'^2}{2|\mb k-\mb p'|^3|\mb k+\mb p'|^3}\Big[2k^2\log^2(-k\tau)-\FR{1}{2}(k^2-p'^2)\log(-k\tau)\Big].
\end{align}
The integral is divergent as $\mb p'\to \pm\mb k$, or equivalently when $p,q\to 0$, and this divergence is removed by the comoving IR cutoff $\LIR$, so we integrate $p$ from $0$ to $k-\LIR$. Now we are ready to carry out the momentum integral, and the final result for the sum of the four diagrams is,
\begin{align}
\label{Pho1L3}
&-\ii\de_{\text{1-loop}}\Big[\wh{G}^{(V)}_{-+}\Big]_{ij}(\mb k,\tau,\tau)\Big|_{\text{scalar,3p}}\n\\
=&~\FR{e^2}{24(2\pi)^2k}\eta_{ij}\Big[\log\Big(\FR{k}{\LIR}\Big)\log^2(-k\tau)+\FR{3\pi-10}{4}\log(-k\tau)\Big]\n\\
=&~\FR{e^2}{24(2\pi)^2k}\eta_{ij}\Big[\log^3(-k\tau)+\FR{3H^2}{2m^2}\log^2(-k\tau)+\FR{3\pi-10}{4}\log(-k\tau)\Big],
\end{align}
where we have used the relation (\ref{LIRm}) in the last line.

\section{Resummation of Late-Time Divergence}
\label{sec5}

In the previous section we calculate the 1-loop corrected in-in propagators of massless scalar, spinor, and vector fields, respectively, to the leading order in late-time divergences. Now we summarize these results here. Firstly, for a scalar field with $\lam\phi^4$ self-interaction, the tree level $-+$ propagator plus 1-loop correction can be got from (\ref{ScaMPP}) and (\ref{Sca1L}) as,
\bge
-\ii\big(1+\de_{\text{1-loop}}\big)\wh{G}_{-+}(\mb k,\tau,\tau)= \FR{H^2}{2k^3}\bigg[1+\FR{\lam H^2}{2(2\pi)^2m^2}\log(-k\tau)\bigg],
\ede
where we have set the renormalization scale $\mu=H$.
Next, the 1-loop corrected propagator of massless Dirac spinor from a massless scalar loop via Yukawa interaction is the sum of (\ref{FerMPP}) and (\ref{Fer1L}) and is given by,
\bge
-\ii\big(1+\de_{\text{1-loop}}\big)\wh{G}^{(F)}_{-+}(\mb k,\tau,\tau)=\FR{H^3\tau^3}{2k}\bigg[k\ga^0-\bigg(1+\FR{y^2}{(2\pi)^2}\log^3(-k\tau)\bigg)\mb k\cdot\bm{\ga}\bigg],
\ede
Finally the 1-loop corrected propagator of the photon from a complex massless scalar field via gauge interaction is the sum of (\ref{PhoMPP}), (\ref{Pho1L4}), and (\ref{Pho1L3}), and the result is,
\bge
-\ii\big(1+\de_{\text{1-loop}}\big)\Big[\wh{G}^{(V)}_{-+}\Big]_{\mu\nu}(\mb k,\tau,\tau)=\FR{1}{2k}\bigg[1+\FR{6e^2H^2}{(2\pi)^2m^2}\log(-k\tau)\bigg]\eta_{\mu\nu}+\FR{e^2}{24(2\pi)^2k}\log^3(-k\tau)\ob{\eta}_{\mu\nu},
\ede
where $\ob{\eta}_{\mu\nu}=\text{diag}(0,1,1,1)$ is the spatial part of flat metric. In all expressions we only keep the leading divergent terms in late-time cutoff $\tau$.

It is clear that all three types of propagators exhibit late-time divergence once 1-loop corrections are included. However, at least in $\lambda\phi^4$ theory of scalar fields, we are quite sure that the late-time divergences presented above at 1-loop level is an artefact of perturbative expansion. As we go to higher order corrections in loop expansion, the degree of late-time divergence, i.e. the power of logarithmic term $\log(-k\tau)$, also increases. Therefore one can imagine that the final result of 2-point function would be convergent as $\tau\to 0$ if we can properly resum the late-time divergent diagrams to all orders in loop expansion. This can be demonstrated exactly in the large-$N$ limit, while for more general cases (including the $\lam\phi^4$ theory of a single real scalar considered here) we can only partially resum the higher order terms. The result of this partial resummation is also convergent as $\tau\to 0$, but it may possess some $\order{1}$ uncertainty as can be recognized by comparing results from different approaches.

With above point clarified, we will now resum the divergences using Dynamical Renormalization Group (DRG) method. The basic idea of DRG is to recognize that the large logs in late time limit is an artefact of perturbation expansion, much like the large logs of scattering energy in loop-corrected scattering amplitude when we extrapolate the loop correction calculated at a given scale to another very different energy scale. The standard way to deal with this problem is to absorb the energy dependence into the running coupling, which is known as the renormalization group (RG) resummation. The resummed result can then be applied to a much wider range of energies, as long as the couplings remain small. The DRG is simply to repeat the same manipulation of RG resummation, replacing large logs of energies by large logs of late times.

  We refer the reader to \cite{Burge09} for a more thorough introduction of DRG method, and here we only point out that for our case, the resummation via DRG is simply to identify the tree-level and 1-loop correction as the first two terms of Taylor expansion of the exponential function. So the resummation goes like $1+f(\tau)\To e^{f(\tau)}$.

The DRG method is conceptually clear and plausible, however, it is not straightforward to understand its proposal diagrammatically, i.e. the resummation via exponentiation. In particular, it is not evident that what loop diagrams are being summed in DRG. To understand this point, suppose that the 1-loop correction has the form of $G_0 f(\tau)$ where $G_0$ is the tree-level result and $f(\tau)$ is the $\tau$-dependent factor from 1-loop correction, then naively we may expect that the resummation is a geometric series,
\bge
\label{NResum}
  G_0\sum_{n=0}^\infty f^n(\tau)=\FR{G_0}{1-f(\tau)}.
\ede
To understand why this is not the case, let us consider a simplified example of mass insertion. The mass term $-\frac{1}{2}m^2\sqrt{-g}\phi^2$ of a real scalar belongs to the ``dangerous interaction'' in the sense that it would lead to late-time divergence once we insert a single mass vertex into the scalar propagator, as is clear from (\ref{massct}). However, this divergence is purely artefact of perturbative expansion, and after resumming all similar mass-insertions, we should be able to recover the divergent free massive propagator built from massive mode function (\ref{MSM}). Now, the naive resummation (\ref{NResum}) can be diagrammatically represented as,
\begin{center}
\includegraphics[width=0.8\textwidth]{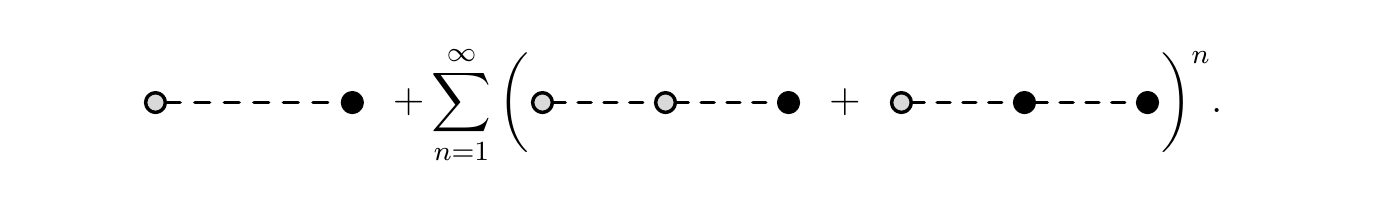}
\end{center}
But this is not the correct expression, because, for instantce, at the level of two mass insertions,
\begin{center}
\includegraphics[width=0.99\textwidth]{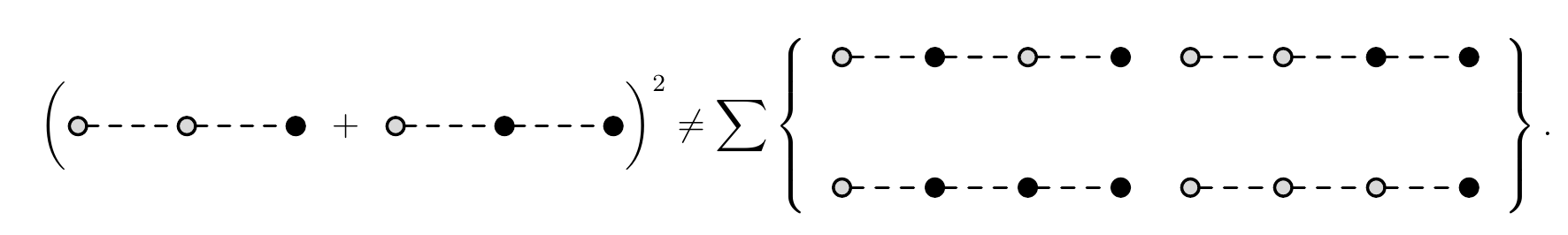}
\end{center}
In above expression, the left hand side is what we would infer from (\ref{NResum}), while the right hand side is the correct diagrams at two-mass-insertion level. Now we are going to show that the resummation via exponentiation suggested by DRG method actually corresponds to resumming all ``correct'' diagrams like the right hand side of above expression, i.e., diagrams with all possible $\pm$ type mass insertions. To this end, we may calculate all mass-insertion diagrams directly and sum them up, but there is a much simpler way, due to the following useful formula,
\begin{align}
\label{MassInsSum}
  &\sum_{a_1,\cdots, a_{n}}\int\FR{\di\tau_1\cdots\di\tau_n}{H^{4n}(\tau_1\cdots\tau_n)^4}a_1\cdots a_n\wh{G}_{-a_1}(\mb k;\tau,\tau_1)\wh{G}_{a_1a_2}(\mb k;\tau_1,\tau_2)\cdots \wh{G}_{a_{n-1}a_n}(\mb k;\tau_{n-1},\tau_n)\wh{G}_{a_n +}(\mb k;\tau_n,\tau)\n\\
  & = \FR{1}{n!}\FR{\pd^n}{\pd(m^2)^n}\wh{G}_{-+}(\mb k;\tau,\tau),
\end{align}
where $a_1,\cdots,a_n=\pm$ denote two types of mass insertions and $\wh G_{\pm\pm}$ is the propagator for 
the scalar field of mass $m$. The above formula is most easily proved in Euclidean de Sitter space, and we refer the readers to \cite{Marolf1006} for details. Here we apply (\ref{MassInsSum}) to a massless scalar propagator and keep the leading divergence in $\tau$ only, then we have the following result,
\bge
  \sum\big(\text{massless propagator with $n$ mass insertions}\big)=\FR{2H^2}{k^3}\FR{1}{n!}\bigg[\FR{2m^2}{3H^2}\log(-k\tau)\bigg]^n
\ede
Finally we sum over $n$, and the exponentiation appears natually,
\bge
  \FR{2H^2}{k^3}(-k\tau)^{2m^2/3H^2}.
\ede
We recognize the above result as the $m/H\ll 1$ limit of the full massive propagator (\ref{MSM}). Therefore, we have justified that the resummation via exponentiation in the case of mass insertion, and showed that the DRG method gives the correct result.

For loop diagrams the situation is more complicated. However, at least for the ``mass-insertion-like'' diagrams like Fig.\,\ref{fig_phi4loop}, it is now clear that the DRG method can resum all such diagrams with all possible $\pm$ vertices, and generates an effective mass to the originally massless propagator as we shall show soon. Here we only note that DRG does not resum more complicated higher loop diagrams, such as sunset diagram at two-loop level, and it is interesting to study how these diagrams would modify the DRG result. For our purpose, it suffices to note that these diagrams would be suppressed by $1/N$ in the large-$N$ limit where $N$ is the number of real scalar fields, and for SM Higgs sector, $N=4$, so DRG result should be a good approximation.

Now we apply DRG resummation to the three cases listed at the beginning of this section. Firstly for scalar field we have,
\bge
-\ii\wh{G}_{-+}(\mb k,\tau,\tau)\Big|_r=\FR{H^2}{2k^3}(-k\tau)^{\lam H^2/2(2\pi m)^2}.
\ede
Then the mass of the scalar field $m$ can be found by comparing the time dependence of the above expression and the propagator of a massive scalar field in the $\tau\to 0$ limit, which is given by $(-\tau)^{-2\sqrt{9/4-(m/H)^2}+3}\simeq (-\tau)^{2m^2/(3H^2)}$, therefore we have,
\bge
\label{ScaMass}
  m^2=\FR{\sqrt{3\lam}H^2}{4\pi}.
\ede
For Dirac fermion we have,
\bge
  -\ii\wh{G}^{(F)}_{-+}(\mb k,\tau,\tau)\Big|_r=\FR{H^3\tau^3}{2k}\bigg[k\ga^0-\exp\bigg(\FR{2y^2}{(2\pi)^2}\log^3(-k\tau)\bigg)\mb k\cdot\bm{\ga}\bigg].
\ede
Here the $k\gamma^0$ term dominate in the $\tau\rightarrow 0$ limit.
As a result, the resummed propogator does not contribute to a mass term for the fermions. Qualitatively, we can understand this conclusion from the mean-field approach. The fermion mass is obtained through its Yukawa coupling $\phi \bar\psi\psi$, which is linear in the Higgs field $\phi$. Thus the mass correction is $m_\mathrm{eff}\sim \langle \phi \rangle = 0$.

Having that said, we should also note that ultimately we are interested in the physical observables, for example, the non-Gaussianity of curvature perturbation induced by fermions. The fermions have to appear in pairs forming closed loops, should they not appear in external lines. Those composite operators such as $\bar\psi \psi$ may obtain non-trivial anomalous dimensions coming from the IR growth of $\phi$. We hope to address this issue in the future.

Finally for the photon, we need to resum the temporal and spatial components separately. For the temporal component we have,
\bge
\label{PRSMT}
  -\ii\Big[\wh{G}_{-+}^{(V)}\Big]_{00}(\mb k,\tau,\tau)\Big|_r=\FR{1}{2k}(-k\tau)^{6e^2H^2/(2\pi m)^2}\eta_{00},
\ede
and for the spatial components we have,
\bge
\label{PRSMS}
-\ii\Big[\wh{G}_{-+}^{(V)}\Big]_{ij}(\mb k,\tau,\tau)\Big|_r=\FR{1}{2k}\exp\bigg(\FR{e^2}{12(2\pi)^2}\log^3(-k\tau)\bigg)\eta_{ij}.
\ede
In expression above, (\ref{PRSMT}) has the form of photon mass contribution while (\ref{PRSMS}) does not. As we can see from (\ref{MPM}), a massive photon's propagator behaves like $(-\tau)^{-2\sqrt{1/4-(M/H)^2}+1}\sim (-\tau)^{2M^2/H^2}$, then comparing this with the above expression we get,
\bge
  M^2=\FR{3e^2H^4}{(2\pi m)^2}=\FR{\sqrt{3}e^2H^2}{\pi\sqrt{\lam}},
\ede
where the first equality agrees with the result in \cite{Proko03} and in the second equality we have used (\ref{ScaMass}). Note that in above expansion of massive photon mode function we have assumed that $M\ll H$, thus the above expression holds only when $e^2/\sqrt{\lam}\ll 1$. If $e^2/\sqrt{\lam}\sim\order{1}$, we should solve the full equation $1-\sqrt{1-(2M/H)^2}=6e^2H^2/(2\pi m)^2$. In particular, if the right hand side of this equation is larger than 1, there will be no real solution to $M$, which means that even (\ref{PRSMT}) cannot be interpreted as an effective mass.

It is worth noting that the convergence of resummed propagators relies crucially on the sign of 1-loop correction. When the 1-loop correction to propagator diverges like $\log^n(-k\tau)$ with $n$ a positive and odd integer at the leading order in $\tau$ (in our situation we always get $n=1$ or 3), the coefficient of this logarithmical factor must be the same with tree-level propagator, so that when $\tau\to 0$, the logarithmic factor goes to $-\infty$, and the resummed factor $\propto \exp [\log^n(-k\tau)]$ is convergent. Were this not the case, the resummed propagator would be even more divergent than the pure 1-loop result. Fortunately in our calculation, all 1-loop corrections have the correct sign to guarantee the convergence of resummation.

\section{Discussions}
\label{sec6}

We have calculated the one-loop corrections of the particle physics Standard Model fields during inflation. Especially, we have calculated the IR scaling behavior of those fields, and such scaling behavior can be interpreted as some sort of mass spectrum for Standard Model fields during inflation.

Rich phenomenology can be derived from the inflationary mass spectrum of the particle physics Standard Model. Especially, the self-coupling parameters of the Standard Model fields are much greater than that of the inflaton. In other words, the fluctuation of the Standard Model sector is non-Gaussian. The non-Gaussianity of the Standard Model fields induces the non-Gaussianity of the density fluctuation. This is because the inflaton has to be coupled to the Standard Model fields with a coupling stronger than gravitational, to allow efficient reheating. The framework to study such non-Gaussianity of density fluctuations is quasi-single field inflation \cite{Chen:2009we,Chen:2009zp,AHM,Baumann:2011nk,Assassi:2012zq,Noumi:2012vr}. We hope to study the relevant signatures in a future work \cite{prep}.

We have relied on the dynamical RG method to resum the diverging contributions and extract the mass correction. The dynamical RG method does not resum all diagrams, but only the reduciable ones. Different approaches have been discussed in the literature by considering the leading logarithm contributions \cite{Starobinsky:1994bd, Bartolo:2007ti, Tsamis:2010iw}. It is not straightforward to extract the mass information from those leading logarithm approaches. However, it remains interesting to understand whether those approaches can improve the knowledge that we obtain from the dynamical RG.

We also note that, once we extrapolate our result to the range $m>3H/2$ for scalar and $m>H/2$ for vector, the loop correction can no longer be interpreted as a mass because the tree-level mass can now introduce non-analytic powers to the mode function, which are not observed in our calculation. We are currently not sure whether this is a fundamental feature of loop corrections, or due to the limitation of the perturbative approach and the resummation scheme. We hope to study this problem in the future.

\paragraph{Acknowledgement.}
We thank Juan Maldacena for invaluable discussions. XC is supported in part by the NSF grant PHY-1417421. YW is supported by the CRF Grants of the Government of the Hong Kong SAR under HKUST4/CRF/13G. ZZX is supported in part by the International Postdoctoral Exchange Fellowship Program of China Postdoctoral Council.

\end{document}